\documentclass[10pt,conference]{IEEEtran}

\usepackage{amsmath,amsfonts}
\usepackage{algorithmic}
\usepackage{graphicx}
\usepackage{textcomp}
\usepackage{subcaption}
\usepackage[utf8]{inputenc}
\usepackage[T1]{fontenc}
\usepackage{hyperref}
\usepackage{url}
\usepackage{nicefrac}
\usepackage{makecell}
\usepackage{multirow}
\usepackage{booktabs}
\usepackage{array}
\usepackage{adjustbox}

\usepackage{listings}
\usepackage{xcolor}
\usepackage{fancyvrb}
\usepackage{tikz}
\usepackage{placeins}
\usepackage{balance}
\usepackage{mdframed}
\usepackage{float}
\definecolor{diffstart}{named}{gray}
\definecolor{diffincl}{RGB}{0,128,0}
\definecolor{diffrem}{RGB}{128,0,0}

\lstdefinelanguage{diff}{
  morecomment=[f][\color{diffstart}]{@@},
  morecomment=[f][\color{diffincl}]{+},
  morecomment=[f][\color{diffrem}]{-},
  morecomment=[f][\color{diffstart}]{---},
  morecomment=[f][\color{diffstart}]{+++},
}

\begin{document}

\title{SWE-Sharp-Bench: A Reproducible Benchmark for C\# Software Engineering Tasks}

\author{
\IEEEauthorblockN{Sanket Mhatre\textsuperscript{*}, Yasharth Bajpai\textsuperscript{*}}
\IEEEauthorblockA{
\textit{Microsoft} \\
Bengaluru, India \\
\{t-smhatre, ybajpai\}@microsoft.com} \\
\and
\IEEEauthorblockN{Sumit Gulwani, Emerson Murphy-Hill, Gustavo Soares}
\IEEEauthorblockA{
\textit{Microsoft} \\
Redmond, WA, USA \\
\{sumitg, emerson.rex, gustavo.soares\}@microsoft.com}
}

\maketitle
\renewcommand{\thefootnote}{\fnsymbol{footnote}}
\footnotetext[1]{Equal Contribution.}
\renewcommand{\thefootnote}{\arabic{footnote}}

\begin{abstract}
AI coding agents have shown great progress on Python software engineering benchmarks 
like SWE-Bench, and for other languages like Java and C in benchmarks like Multi-SWE-Bench.
However, C\# -- a prominent enterprise language ranking \#5 in the TIOBE index -- 
remains absent from such benchmarks. 
We introduce SWE-Sharp-Bench, 
a reproducible software engineering benchmark for C\# featuring 150 instances from 17 repositories. 
Evaluating identical model-agent configurations across languages reveals a significant performance gap:
while 70\% of Python tasks in SWE-Bench Verified are solved, 
only 40\% of our C\# tasks are resolved. 
We open-source SWE-Sharp-Bench and our entire curation pipeline.
\end{abstract}


\newcommand{\todo}[1]{\textcolor{red}{\textbf{ #1\textbf{]]}}}}

\newcommand{\ysrt}[1]{{\todo{Yasharth:  {\color{green} #1}}}}
\newcommand{\bhavya}[1]{{\todo{Sanket:  {\color{cyan} #1}}}}
\newcommand{\gustavo}[1]{{\todo{Gustavo:  {\color{blue} #1}}}}
\newcommand{\sumit}[1]{{\todo{Sumit:  {\color{purple} #1}}}}

\section{Introduction}  
Large Language Model-powered automated software engineering has gained 
substantial attention due to its potential for increasing developer productivity. 
These models now enable code completion, unit test generation, documentation generation, 
interactive chat interfaces, and -- more recently -- autonomous coding agents~\cite{10.5555/3737916.3739517, 10714559}. 
This has necessitated increasingly sophisticated benchmarks, moving from simple function-level code completion to 
evaluating these models on their ability to autonomously solve real-world software engineering problems.
SWE-Bench~\cite{jimenez2024swebench} and SWE-Bench Verified~\cite{chowdhury2024swebenchverified} 
have become the most widely-used benchmarks for evaluating latest models and agents on software engineering tasks.
Other software engineering benchmarks -- Multi-SWE-Bench~\cite{zan2025multiswebenchmultilingualbenchmarkissue}, 
SWE-Bench Multilingual, and SWE-PolyBench -- cover additional programming languages beyond Python. 

However, examining the TIOBE Programming Community Index -- ``an indicator of the popularity of programming languages''\footnote{\url{https://www.tiobe.com/tiobe-index/}} -- reveals systematic gaps in benchmark coverage. 
The top 8 languages by popularity are: 
Python (\#1, covered by SWE-Bench), 
C++ (\#2, covered by Multi-SWE-Bench), 
C (\#3, covered by Multi-SWE-Bench),
Java (\#4, covered by Multi-SWE-Bench and SWE-PolyBench), 
C\# (\#5, \textbf{no coverage}), 
JavaScript (\#6, covered by Multi-SWE-Bench and SWE-PolyBench), 
Visual Basic (\#7, \textbf{no coverage}), and
Go (\#8, covered by Multi-SWE-Bench).
Notably, the entire .NET ecosystem -- including both C\# and Visual Basic -- remains absent from
software engineering benchmarks, despite their high rankings. This gap is particularly striking given C\#'s importance to enterprise software development.\footnote{\url{https://dotnet.microsoft.com/en-us/platform/customers}}
The absence of .NET languages limits our understanding of how models and coding agents
perform with C\#'s unique characteristics.

In this paper, we introduce SWE-Sharp-Bench, the first software engineering task benchmark 
for the C\# and .NET ecosystem, comprising 150 curated instances, by adapting SWE-Bench's methodology. 
Our curation pipeline tackles .NET-specific challenges including sophisticated dependency management, 
multi-version compatibility, and cross-platform development to ensure automated creation of reproducible 
containerized environments. Evaluating these instances on leading models from OpenAI and Anthropic 
using popular agent frameworks (SWE-Agent and OpenHands~\cite{wang2025openhands}), 
we reveal that C\# presents significant challenges for current models, with performance gaps 
that appear to stem from the relatively high complexity of typical changes in C\# projects.

\begin{figure*}[!ht]
    \centering
    \includegraphics[width=1\textwidth]{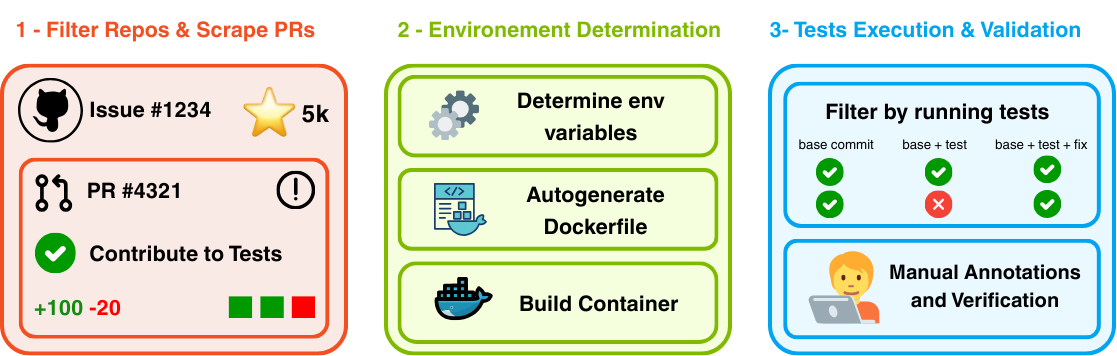}
    \caption{Curation pipeline}
    \label{fig:placeholder}
\end{figure*}

\section{Related Work }  

Early code-generation benchmarks such as HumanEval \cite{chen2021evaluatinglargelanguagemodels} and MBPP \cite{austin2021programsynthesislargelanguage} established the standard for code-generation evaluation. This approach was subsequently scaled and generalized in HumanEval-XL, MBXP, and MultiPL-E \cite{10.1109/TSE.2023.3267446} (which includes C\#). However, these benchmarks still primarily evaluated small, self-contained programming tasks. A shift toward repository-level software-engineering evaluation emerged with SWE-Bench and SWE-Bench Verified, which assess real-world pull requests from open-source repositories. This methodology has since broadened along several axes. In the multilingual direction, Multi-SWE-Bench provides 1,632 instances across seven languages (Java, JavaScript, Go, Rust, C, and C++), SWE-Bench Multilingual\footnote{\url{https://kabirk.com/multilingual}} contributes 300 instances spanning nine languages (C, C++, Java, JavaScript, Go, Rust, TypeScript, PHP, and Ruby), and SWE-PolyBench focuses on JavaScript, TypeScript, Python, and Java. In parallel, GitBug-Java \cite{Saavedra_2024} develops a similar repository-level benchmark for Java, using GitHub Actions to ensure reproducible builds. In addition to text-only contexts, SWE-Bench Multimodal \cite{yang2025swebench} extends evaluation to issues that require visual understanding. Beyond curated issues from open-source repositories, SWE-Lancer \cite{miserendino2025swelancer} introduces end-to-end tasks sourced from freelancing platforms. Recent work, such as SWE-Smith \cite{yang2025swesmithscalingdatasoftware}, also explore scaling instance creation via synthetic data generation with LLMs. Despite these multilingual and methodological advances, C\# remains underrepresented in repository-level software-engineering evaluation.

\section{Building SWE-Sharp-Bench}  

\begin{figure*}
    \centering
    \includegraphics[width=1\textwidth]{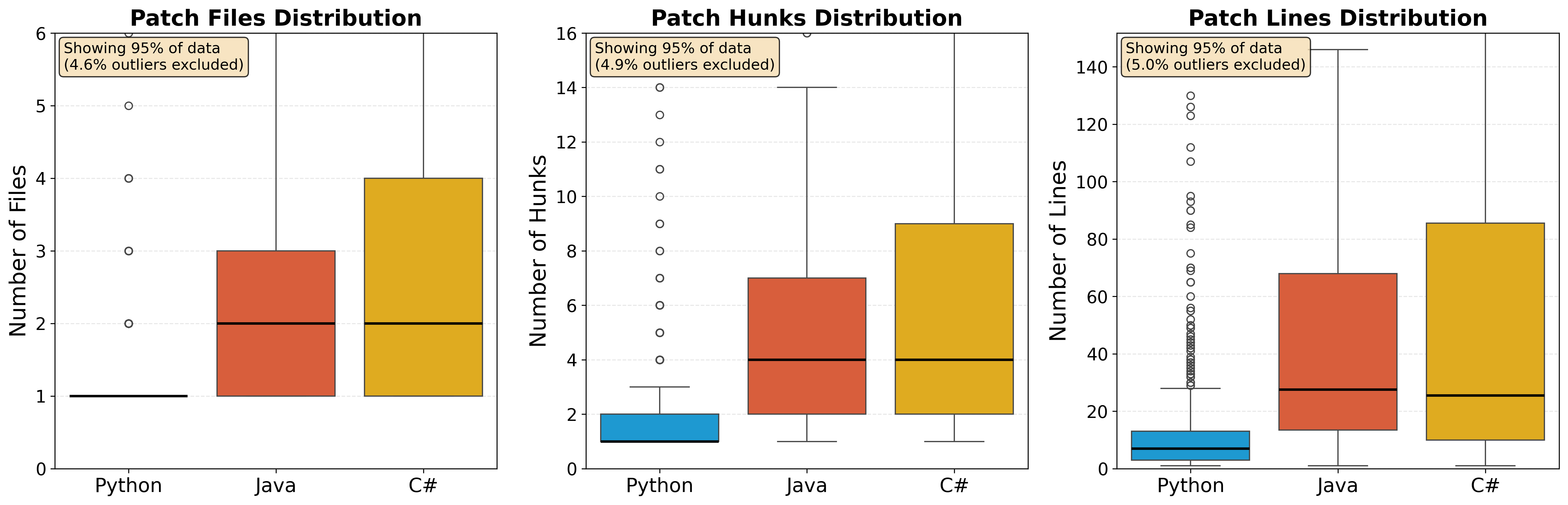}
    \caption{Distribution of \#Files, \#Hunks and \#Lines in Patches across Languages}
    \label{fig:violin_plots_by_language}
\end{figure*}

SWE-Sharp-Bench consists of 150 issue-resolving tasks carefully curated from 17 popular and actively maintained C\# GitHub repositories. The tasks are mapped to GitHub issues which either reports bugs or requests a new feature. 

\subsection{Benchmark Construction}
\textbf{1. Repository Selection:} From the top 100 C\# GitHub repositories, we retain projects with at least 5,000 stars, active maintenance in the past 6 months, and verified build viability.\\
\textbf{2. PR Scraping \& Attribute Filtering:} We scrape the 1,000 most recent PRs per selected repository and retain only that (1) reference at least one GitHub issue, (2) modify test files and (3) were successfully merged into repository's default branch.\\
\textbf{3. Environment Determination:} For each PR, we auto-generate a Dockerfile by parsing \texttt{.github/workflows}, \texttt{*.sln}, \texttt{*.csproj}, \texttt{global.json} and \texttt{.env} to infer NuGet dependencies, build targets and environment variables. The generator reconciles version declarations and handles .NET-specific hurdles (NuGet/MSBuild asset selection, multi-targeting conflicts, deprecated runtimes, etc). We validate by building and running the container.\\
\textbf{4. Execution-based Filtering:} For each PR, we run tests at (1) base , (2) base + test patch and (3) base + test + fix, retaining only cases with \textit{pass} $\rightarrow$ \textit{fail} $\rightarrow$ \textit{pass}; all others are treated as flaky and omitted.\\
\textbf{5. Manual Verification:} Each candidate PR was independently annotated and cross-checked by the first two authors with standards similar to SWE-Bench Verified. We flag (1) under specified problem statements and (2) inadequate tests ( overly narrow or misaligned ). Only PRs that pass this review are considered for the final benchmark.

Full details about the benchmark construction process are discussed in Appendix.

\subsection{Benchmark Characterization}

\subsubsection{Features of SWE-Sharp-Bench}
SWE-Sharp-Bench represents a variety of tools and applications. The repositories can be categorized into Data \& Storage (4), API Infrastructure (3), User Interface(3), Development Tools (5) and Multimedia Processing (2). The categorization is done by manual inspection of the repository descriptions (see Appendix Table \ref{tab:repository_descriptions}).
The instances are categorized into three primary categories Bug-Fixes (91), Feature Requests (47) and Others (12). The categorization process is discussed in Appendix \ref{app:benchmark_categorization} .
53\% of the instances are created in 2024 and 90\% of the instances are created after 2023. 
 
\subsubsection{Characteristics of SWE-Sharp-Bench versus Other Benchmarks}
\textbf{\\Language and Benchmark Selection:}
Before attributing performance differences to model or agent limitations, we need to understand the difference in characteristics of the current benchmarks. We select Python, the essential baseline with highest representation in research interests and Java which represents a natural comparison point for C\# as it shares similar properties like static typing, complex build systems and dependency management. We use SWE-Bench-Verified for Python and Multi-SWE-Bench for Java.\\
\textbf{Patch Complexity Analysis:}
We adopt the patch complexity metrics introduced in Multi-SWE-Bench, which measures static properties across three dimensions:\\
$\bullet$ Patch-level metrics: Files modified (change breadth), hunks per patch (modification granularity), lines added/removed (change magnitude). \\
$\bullet$ Repository metrics: Total files and lines of code.\\
$\bullet$ Task specification: Token length of the problem statement.\\
 We extract these metrics for the 150 SWE-Sharp-Bench instances and 500 SWE-Bench Verified instances. For Java, we use the metrics reported in the Multi-SWE-Bench paper. Figure \ref{fig:violin_plots_by_language} demonstrates the distribution of files modified, number of hunks and number of lines added in patches across languages. 
 In the Appendix, Tables \ref{tab:instance_statistics}, \ref{tab:instance_statistics_swe-bench-verified} and \ref{tab:multi_swe_bench_statistics} summarize repository-level statistics. Our analysis using these metrics reveal:\\
\textbf{Repository Scale:} C\# projects range from roughly 5$\text{k}$ LoC to 1.47 $\text{M}$ LoC (median~$\approx235\,\text{}k$), whereas the largest Python repository in SWE‑Bench Verified is 383 $\text{k}$ LoC and the largest Java repository in Multi‑SWE‑Bench reaches 443 $\text{k}$ LoC. Bigger repositories translate to a broader surface area for the agent to search through. \\
\textbf{Change Locality:} Python fixes are usually surgical (mean=1.24, median=1 file); Java shows moderate spread (mean=2.96, median=2); and C\# has the broadest distribution (mean=4.88, median=2), combining many small fixes with a long tail of multi-file changes, creating a diverse mix.\\
\textbf{Modification Granularity:} Patch depth increases progressively from Python (2.4 hunks and 14.3 lines on average ) to Java (6.26/89.27) to C\# (10.0/131.1), as shown in Figure~\ref{fig:violin_plots_by_language}, middle. C\# exhibits the most diverse distribution: like Python and Java, many patches are small, but C\# also includes substantially larger modifications. Manual inspection of high patch-count C\# instances revealed these typically involve refactoring operations and coordinated multi-file edits.\\
\textbf{Task Specification:} Problem‑statement length varies markedly by language. Java issues can exceed $\sim$1.6 $\text{k}$ tokens, Python issues generally stay below 450, and C\# issues are often under 150. 
Patches with short descriptions may be difficult for agents to solve if they are insufficiently detailed.\\
\textbf{Takeaways:} C\# exhibits the most complex static patch properties among the three languages, and SWE-Sharp-Bench provides a diverse mix of task complexities.

\section{Experiments and Results}

\begin{table}[tb]
\centering
\caption{Resolution Rates (\%) of Open-Source agents across various models on SWE-Sharp-Bench}
\label{tab:resolve_rate_open_agents}
\begin{tabular}{@{}lcc@{}}
\toprule
\textbf{Model} & \textbf{SWE-Agent} & \textbf{OpenHands} \\
\midrule
GPT-4o             & 11.3 & 8.0  \\
GPT-4.1            & 22.0 & 23.3 \\
GPT-5              & 43.3 & 47.3 \\
\midrule
o3-mini     & 19.3 & 19.3 \\
o4-mini     & 25.3 & 26.0 \\
o3          & 33.3 & 35.0 \\
\midrule
Claude Sonnet 3.5  & 20.0 & 22.6 \\
Claude Sonnet 3.7  & 31.3 & 31.3 \\
Claude Sonnet 4    & 44.7 & 40.6 \\
\bottomrule
\end{tabular}
\end{table}

\textbf{Agents \& Models:}
We evaluate two popular agent systems: SWE-Agent and OpenHands. We test each system across multiple leading language models from OpenAI and Anthropic. Since these frameworks were originally designed for Python repositories, we adapt their prompts for C\# projects. Each agent receives a single attempt per instance with a 2-hour timeout limit. due to budget constraints, we conduct a single attempt per instance, retrying only when infrastructure failures (e.g., rate limits, API errors ) occur to ensure at least one valid attempt per instance.\\

\textbf{Evaluation Metrics:}
We use \textbf{Resolution rate} as the primary metric, the percentage of instances successfully resolved by each agent. An instance is considered resolved when the agent's generated patch passes all the required tests.\\
\textbf{Results:} Table \ref{tab:resolve_rate_open_agents} summarizes performance on SWE-Sharp-Bench of different OpenAI and Anthropic models using SWE-Agent and OpenHands. Across all configurations, OpenHands + GPT-5 performs the best with 47.3 \% resolution rate. Table \ref{tab:benchmark_results} in Appendix, demonstrates performance on SWE-Bench Verified and Multi-SWE-Bench' Java subset on identical model-agent combinations. More results are discussed in the Appendix.

\begin{figure}
    \centering
    \includegraphics[width=0.5\textwidth]{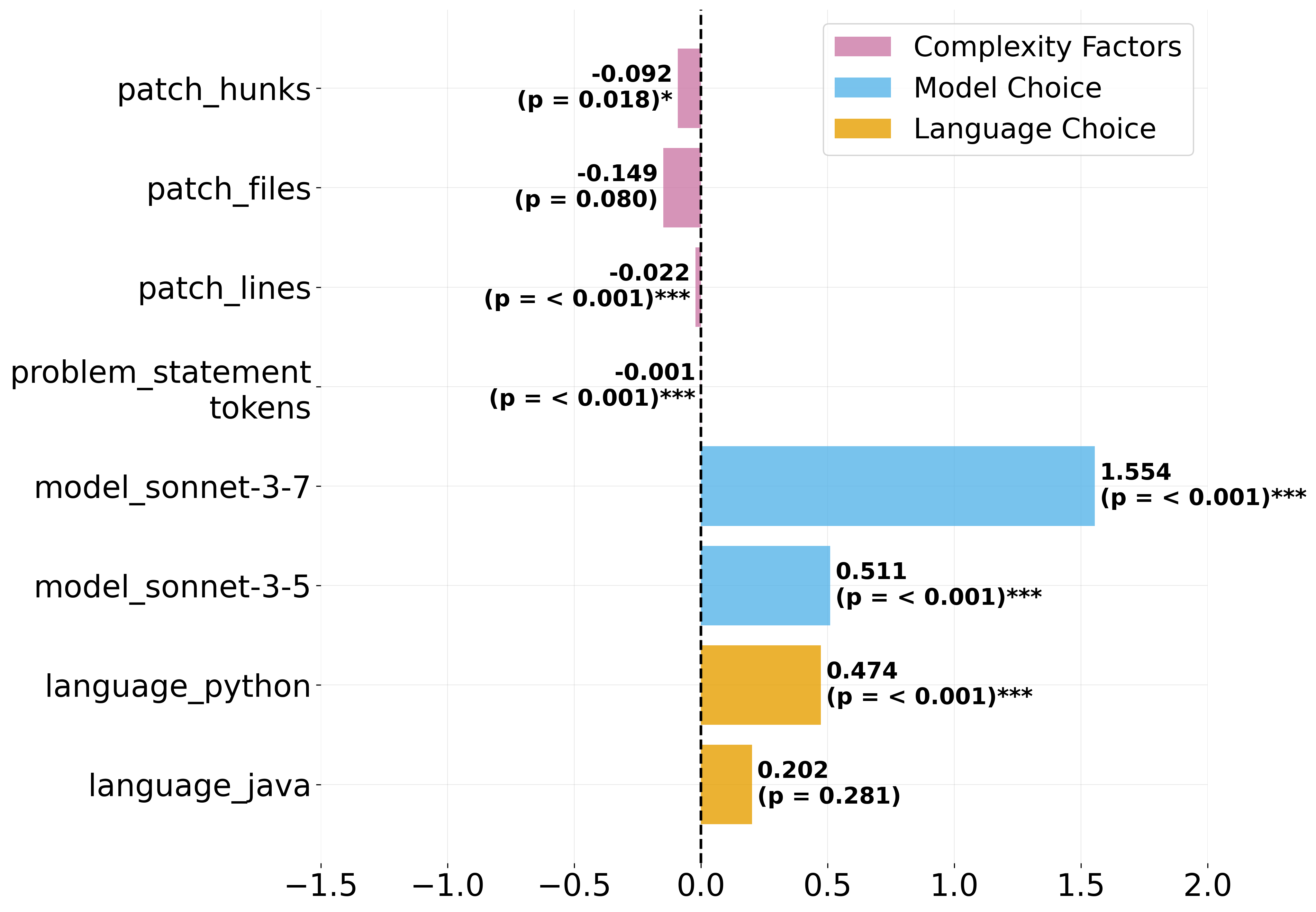}
    
    \caption{Logistic regression coefficients and their influence on resolution rate (baseline: C\#, GPT-4o). Significance: *** p < 0.001, ** p < 0.01, * p < 0.05.}
    \label{fig:logistic_regression}
\end{figure}

\subsection{Which factors affect agent performance ?}
We observe performance gaps between Python when compared to C\# and Java for identical model-agent configurations, e.g. on SWE-Agent + Claude Sonnet 3.7 , Resolution rate for Python is 62.40 \% , 30.67 \% for C\# and 14.68 \% for Java. Refer Table 5 in the appendix for more details. The likelihood of successful instance resolution depends on multiple factors: how spread out the patch is ( hunks and files in the patch), how big the patch is ( patch lines), which model was used, or which agent was used. We observe that agents perform significantly better on Python compared to C\# and Java, but it's unclear how much of this performance gap is correlated with the instance's static properties versus the choice of programming language itself. To understand which factors most strongly influence resolution success, we use logistic regression analysis. We conduct this analysis using instance-level data from SWE-Agent runs with GPT-4o, Claude Sonnet 3.5 and Claude Sonnet 3.7 models, the combinations with complete instance level resolution data available across all three benchmarks. 

Figure \ref{fig:logistic_regression} shows the regression coefficients with GPT-4o and C\# as baseline categories. As expected, model choice has the strongest influence, with newer models significantly outperforming GPT-4o. More complexity decreases success: patch hunks, lines, and files negatively affect performance, indicating that widespread edits across multiple locations are more challenging for agents to resolve. Programming language shows a substantial effect, with Python significantly easier than the C\# baseline. This analysis confirms that Python is the easiest and -- after controlling for complexity and model -- Java and C\# are similarly difficult.

\section{Limitations}
While SWE-Sharp-Bench provides a diverse mix of tasks, it has few limitations. First, with 150 instances, it is smaller than SWE-Bench Verified's 500, though comparable to individual language subsets in Multi-SWE-Bench. Second, unlike SWE-Bench Verified and Multi-SWE-Bench, we do not provide manual annotations for difficulty, though patch complexity metrics offer objective complexity indicators. Finally, as with any benchmark derived from public data, the included data might have been used for any recent LLM training.

\section{Benchmark \& Data Availability}

The Appendix contains detailed analysis, a breakdown of the benchmark, and a deep dive into the performance results. Benchmark data is available at HuggingFace\footnote{\href{https://huggingface.co/datasets/microsoft/SWE-Sharp-Bench}{huggingface.co/datasets/microsoft/SWE-Sharp-Bench}}. The curation pipeline code and agent trajectories are available at GitHub\footnote{\href{https://github.com/microsoft/prose/tree/main/misc/SWE-Sharp-Bench}{github.com/microsoft/prose/tree/main/misc/SWE-Sharp-Bench}}.

\FloatBarrier

\bibliographystyle{IEEEtran}
\bibliography{references}

\clearpage
\clearpage
\appendix

\section{Appendix}

\subsection{C\# Background}
For readers unfamiliar with C\#, understanding its project structure is essential to appreciating the unique challenges it presents for automated environment building. While Python and Java projects can certainly be complex with their own packaging systems, C\# introduces additional layers of hierarchy through its project system, A \texttt{.sln} solution file acts as a master container, similar to a workspace, that can contain multiple \texttt{.csproj} (C\# project) files, each defining a separate component like a main application, test suite or shared library. This multi-layered structure becomes particularly challenging with C\#'s multi-targeting feature, where a single project can be compiled for different .NET versions simultaneously (eg, .NET 6.0, .NET Framework 4.8 and .NET standard 2.0). Imagine if a Python project needed to maintain compatibility with Python2.7, 3.8 and 3.11 simultaneously within the same code base, with different APIs for each version, this is routine in C\#. This architectural complexity, combined with various build configurations, platform targets and the tight coupling between Visual Studio tooling and project files, make C\# repositories more challenging to automatically analyze and test compared to the relatively flat structure of Python packages or Java's more uniform build systems like Maven or Gradle. These intricate inter-dependencies mean that what might be a simple \texttt{pip install} and \texttt{pytest} in Python becomes a complex orchestration of MSBuild targets, NuGet package restoration and framework-specific test runners in C\#. 

\subsection{Benchmark Construction}\label{app:benchmark_construction}

\emph{\textbf{1. Repository Selection:}}
We select a high-quality set of GitHub repositories through a multi-stage filtering process.\\
\textbf{Popularity Filtering}: We identified the top 100 C\# repositories on GitHub ranked by GitHub stars. From this initial set, we applied a minimum threshold of 5,000 stars.\\
\textbf{Active Maintenance}: We verify that selected repositories demonstrate active development and maintenance within the last 6 months by manually analyzing three indicators: commit history frequency, merged pull requests, and issue creation activity.\\
\textbf{Build Viability}: Verification by either minimal manual setup or executing local GitHub Actions workflows with repo's latest commit. Only repositories that successfully build and pass their test suites at the latest commit were considered.

\emph{\textbf{2. Pull Request Scraping and Attribute-based Filtering:}}
We start by scraping the 1000 most recent PRs from each selected repository. All PRs are then filtered by the following criteria:\\
\textbf{Linked with at least one GitHub issue:} The PR must reference at least one GitHub issue to ensure it addresses either a bug report or a feature request.\\ 
\textbf{Changes to test files:} The PR must include modifications to test files, indicating the author contributed to testing to ensure the issue is resolved. We identify test files using keyword pattern matching for "test" or "testing" in filenames or filepaths, which captures the majority of test files in our dataset.\\
\textbf{Merged in main branch:} The PR must be successfully merged into the main branch, indicating the PR was thoroughly reviewed and approved by repository maintainers.

\emph{\textbf{3. Environment Determination: }}
To ensure consistently reproducible environments, for each PR we automatically construct a Dockerfile which parses .github/workflows, .sln, .env, global.json and .csproj files to automatically determine the project's environment variables and dependencies. This automated construction addresses several unique challenges in .NET containerization: managing complex .NET dependency resolution compared to simpler package managers like pip, handling multiple .NET framework versions within single containers, dealing with deprecated .NET versions that require specific base images, and resolving projects that target multiple .NET versions simultaneously. Our system automatically detects environment variables from various configuration sources and reconciles version conflicts across different project files.

We validate each Dockerfile by creating the Docker image and launching the container to ensure consistent environment setup. Any failed builds are manually analyzed for missing dependencies, misconfigurations, or version conflicts. Such issues are fixed if they require minimal modifications to the Dockerfile that do not involve significant human effort. Otherwise, such PRs are discarded from our dataset.

\emph{\textbf{4. Execution-based Filtering: }}
For each PR, we run tests in three states: at the base commit, after applying the test patch at the base commit and finally applying both fix and test patch at the base commit. We filter PRs to include only those demonstrate this pattern: all test pass at the base commit, at least one test fails after applying the test patch and again all tests pass when we apply both fix and test patch. Any other scenarios are marked flaky and discarded from the dataset.

Implementing this execution-based validation for .NET projects required addressing different challenges compared to ecosystems like Python. Unlike Python's widely used and more standardized pytest ecosystem, different .NET projects utilize different testing frameworks (NUnit, XUnit, MSTest) each with different execution patterns and output formats. Our system automatically detects and handles these testing frameworks, unifies their logging mechanisms and normalizes their output formats to support a common evaluation harness across all repos. 

\emph{\textbf{5. Manual Verification: }}
Finally, each PR is annotated and cross-verified by first two authors. The annotations were carried out with standards set similar to SWE-Bench Verified. The candidate PRs are checked for:\\
\textbf{Underspecified Problem Statements:} We annotate if the problem statement is underspecified, leading to ambiguity on what the problem is or how it should be solved.\\
\textbf{Thorough Test Cases:} We check whether the test cases introduced in the PR are overly specific, narrow, or unrelated to the actual problem being solved.\\
Instances which pass this manual inspection were selected for the final benchmark.

\onecolumn 
\begin{figure*}
    \centering
    \includegraphics[width=1\textwidth]{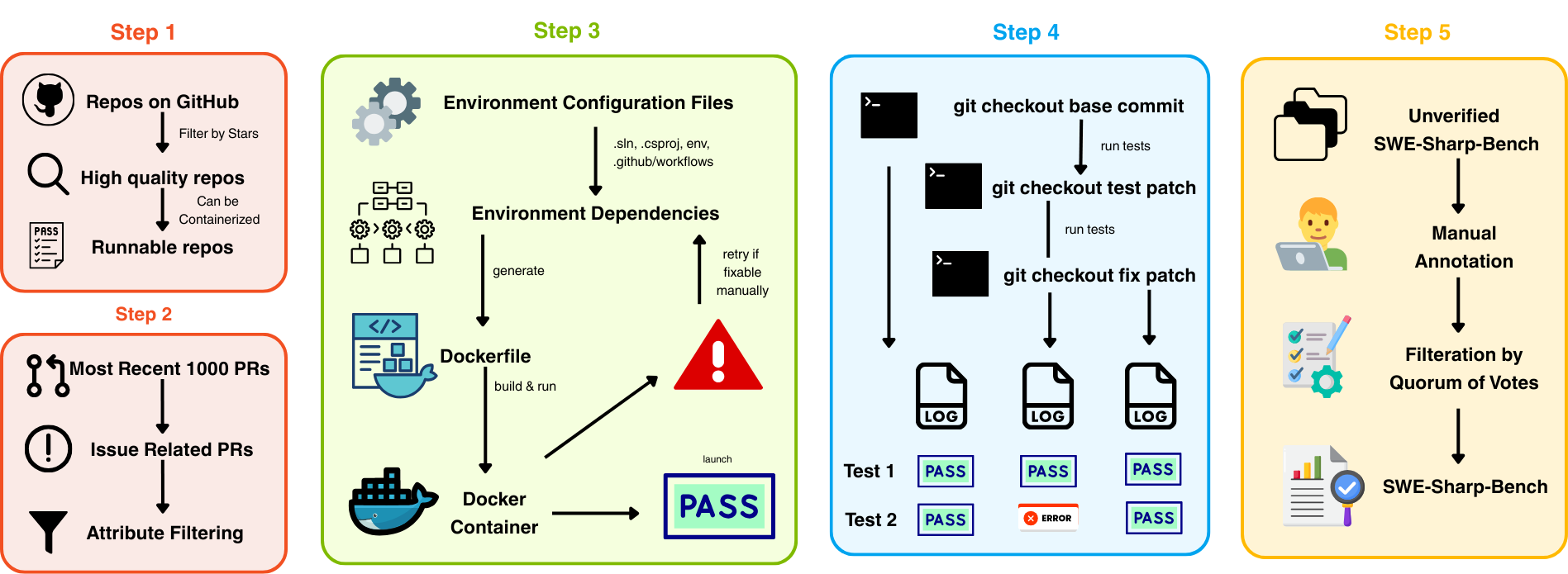}
    \caption{SWE-Sharp-Bench Curation Process}
    \label{fig:SWE-Sharp-Bench Curation Process}
\end{figure*}

\begin{table*}
    
\center
\caption{Detailed Repository Descriptions}
\label{tab:repository_descriptions}

\begin{tabular}{|p{4.5cm}|p{9.5cm}|}
\hline
\textbf{Repository} & \textbf{Description} \\
\hline
App-vNext/Polly & Polly is a .NET resilience and transient-fault-handling library that allows developers to express policies such as Retry, Circuit Breaker, Timeout, Bulkhead Isolation, and Fallback in a fluent and thread-safe manner. \\
\hline
AvaloniaUI/Avalonia & Develop Desktop, Embedded, Mobile and WebAssembly apps with C\# and XAML. The most popular .NET UI client technology \\
\hline
JoshClose/CsvHelper & Library to help reading and writing CSV files \\
\hline
MessagePack-CSharp/MessagePack-CSharp & Extremely Fast MessagePack Serializer for C\#(.NET, .NET Core, Unity, Xamarin). / msgpack.org[C\#] \\
\hline
SixLabors/ImageSharp & A modern, cross-platform, 2D Graphics library for .NET \\
\hline
StackExchange/\newline StackExchange.Redis & General purpose redis client \\
\hline
ThreeMammals/Ocelot & .NET API Gateway \\
\hline
ardalis/CleanArchitecture & Clean Architecture Solution Template: A proven Clean Architecture Template for ASP.NET Core 9 \\
\hline
autofac/Autofac & An addictive .NET IoC container \\
\hline
devlooped/moq & The most popular and friendly mocking framework for .NET \\
\hline
dotnet/BenchmarkDotNet & Powerful .NET library for benchmarking \\
\hline
dotnet/efcore & EF Core is a modern object-database mapper for .NET. It supports LINQ queries, change tracking, updates, and schema migrations. \\
\hline
gui-cs/Terminal.Gui & Cross Platform Terminal UI toolkit for .NET \\
\hline
jellyfin/jellyfin & The Free Software Media System - Server Backend \& API \\
\hline
restsharp/RestSharp & Simple REST and HTTP API Client for .NET \\
\hline
serilog/serilog & Simple .NET logging with fully-structured events \\
\hline
spectreconsole/spectre.console & A .NET library that makes it easier to create beautiful console applications. \\
\hline
\end{tabular}

\end{table*}

\subsection{Benchmark Characterization} \label{app:benchmark_categorization}

\paragraph{\textbf{Issue Type Categorization and Distribution}}
We systematically categorize instances into three primary types: Bug-Fix (97), Feature Requests (41), and Refactors (12), with further subdivision into secondary subtypes. Bug-Fix issues have Logic (54), UI (28), Logging (6), Concurrency (2), Compatibility (2), Stoage (2), API (1), Networking (1) and Security (1). Feature Requests have API (16), Logic (6), UI (6), Configuration (4), Logging (4), Storage (2), Documentation (1), Performance (1) and Compatibility (1) as sub-types. For this categorization, we selected the labels by manual inspection of few sampled tasks. After deciding on the labels, we used GPT-5 to provide a label using problem statement, patch and the repo's description. \ref{fig:characterization_prompt}

\begin{figure*}[!ht]
\begin{mdframed}[linewidth=1pt, roundcorner=5pt]
\begin{lstlisting}[basicstyle=\ttfamily\small, breaklines=true]
TASK: Classify software development issues into categories based on the provided information.

INPUT PROVIDED:
1. Repository Information: Details about the software project
2. Problem Statement: Description of the issue or feature request
3. Code Changes (Patch): The actual code modifications made to address the issue

CLASSIFICATION CATEGORIES:

PRIMARY LABEL (required - select exactly one):
-> Feature: Adding new functionality, enhancements, or feature requests
-> Bug: Fixing incorrect behavior, errors, or defects
-> Other: Documentation updates, build system changes, infrastructure, etc.

SECONDARY LABEL (required - select the most relevant one):
- UI: User interface, visual elements, controls, rendering issues
- Performance: Speed optimization, resource usage improvements
- Security: Authentication, authorization, vulnerability fixes
- Storage: Database operations, file system, data persistence
- Logging: Logging systems, telemetry, monitoring, tracing
- API: External interfaces, endpoints, API integration
- Configuration: Settings, options, setup, environment configuration  
- Testing: Test frameworks, test utilities, validation logic
- Documentation: Documentation updates, comments, help text
- Build: Compilation, packaging, deployment, build system
- Compatibility: Version compatibility, platform support
- Logic: Business logic, algorithms, calculations, core functionality
- Memory: Memory management, memory leaks, allocation issues
- Concurrency: Threading, async operations, race conditions
- Networking: HTTP, TCP, network protocols, connectivity
- Other: If none of the above categories fit well

INSTRUCTIONS:
- Analyze the repository context to understand the project domain
- Read the problem statement to understand what needs to be addressed
- Examine the code changes to see what was actually implemented/fixed
- Choose the PRIMARY label based on whether this adds functionality (Feature), fixes a problem (Bug), or is maintenance work (Other)  
- Choose the SECONDARY label based on the technical area most affected by the changes

OUTPUT FORMAT (JSON only):
{{
"primary": "Feature|Bug|Other",
"secondary": "UI|Performance|Security|Storage|Logging|API|Configuration|Testing|Documentation|Build|Compatibility|Logic|Memory|Concurrency|Networking|Other"
}}

---

{repo_description}

Problem Statement:
{problem_statement}

Code Changes (Patch):
{patch}

Analyze the above information and provide your classification in the exact JSON format specified.
\end{lstlisting}
\caption{Prompt Template used for Issue Type Categorization}
\label{fig:characterization_prompt}
\end{mdframed}
\end{figure*}

\FloatBarrier
\onecolumn 
\subsection{Instance Statistics}
\begin{table*}[!htb]
    \begin{center}
\captionof{table}{Statistics of SWE-Sharp-Bench}
\label{tab:instance_statistics}
\resizebox{\textwidth}{!}{%
\begin{tabular}{l|rr|rr|rrr|rrr|rr}
\toprule
\multirow{2}{*}{\textbf{Org/Repo}} & \multicolumn{2}{c|}{\textbf{Repository}} & \multicolumn{2}{c|}{\textbf{Instance}} & \multicolumn{3}{c|}{\textbf{Gold Patch}} & \multicolumn{3}{c|}{\textbf{Test Patch}} & \multicolumn{2}{c}{\textbf{Unit tests}} \\
& \textbf{\#Files} & \textbf{\#LoC} & \textbf{\#Count} & \textbf{Avg. \#Tokens} & \textbf{Avg. \#Lines} & \textbf{Avg. \#Hunks} & \textbf{Avg. \#Files} & \textbf{Avg. \#Lines} & \textbf{Avg. \#Hunks} & \textbf{Avg. \#Files} & \textbf{\#Avg. F2P} & \textbf{\#Avg. P2P} \\
\midrule
App-vNext/Polly & 790 & 84.8k & 14 & 176.64 & 98.57 & 13.64 & 7.50 & 123.79 & 14.21 & 6.21 & 1.64 & 5.36 \\
AvaloniaUI/Avalonia & 3532 & 436.5k & 41 & 287.02 & 147.95 & 8.80 & 3.73 & 63.51 & 3.54 & 1.32 & 2.29 & 4.10 \\
JoshClose/CsvHelper & 461 & 44.8k & 1 & 192.00 & 369.00 & 35.00 & 22.00 & 423.00 & 36.00 & 30.00 & 13.00 & 10.00 \\
MessagePack-CSharp/MessagePack-CSharp & 688 & 80.8k & 18 & 292.72 & 41.61 & 7.11 & 2.28 & 117.78 & 3.39 & 2.33 & 1.67 & 0.56 \\
SixLabors/ImageSharp & 1997 & 248.4k & 8 & 256.38 & 56.00 & 4.62 & 2.62 & 42.75 & 4.62 & 2.38 & 0.75 & 6.88 \\
StackExchange/StackExchange.Redis & 341 & 72.8k & 3 & 406.67 & 50.67 & 9.67 & 5.00 & 104.33 & 15.33 & 4.67 & 2.67 & 0.00 \\
ThreeMammals/Ocelot & 735 & 53.5k & 4 & 146.50 & 498.75 & 44.00 & 24.50 & 699.50 & 31.50 & 17.25 & 10.25 & 7.50 \\
ardalis/CleanArchitecture & 256 & 5.0k & 3 & 125.33 & 34.33 & 4.33 & 4.33 & 45.00 & 6.67 & 5.67 & 3.33 & 0.33 \\
autofac/Autofac & 577 & 45.8k & 6 & 411.83 & 134.83 & 10.67 & 5.17 & 83.33 & 2.00 & 1.33 & 3.83 & 13.33 \\
devlooped/moq & 242 & 40.8k & 4 & 287.75 & 34.25 & 4.50 & 3.00 & 55.50 & 2.00 & 1.00 & 3.00 & 0.00 \\
dotnet/BenchmarkDotNet & 955 & 67.0k & 8 & 291.88 & 54.75 & 6.00 & 3.38 & 71.88 & 2.12 & 1.12 & 3.62 & 14.00 \\
dotnet/efcore & 5468 & 1479.1k & 9 & 232.78 & 565.00 & 20.11 & 8.22 & 174.11 & 11.22 & 5.89 & 1.00 & 14.22 \\
gui-cs/Terminal.Gui & 938 & 221.7k & 1 & 113.00 & 22.00 & 3.00 & 2.00 & 5.00 & 1.00 & 1.00 & 1.00 & 0.00 \\
jellyfin/jellyfin & 1943 & 224.7k & 3 & 679.00 & 23.00 & 2.00 & 2.00 & 30.33 & 1.33 & 1.00 & 1.00 & 6.67 \\
restsharp/RestSharp & 226 & 15.9k & 5 & 192.00 & 66.60 & 11.60 & 5.20 & 37.40 & 4.40 & 2.40 & 1.80 & 2.40 \\
serilog/serilog & 214 & 21.2k & 12 & 257.08 & 80.50 & 7.42 & 3.25 & 84.25 & 8.08 & 4.92 & 3.25 & 4.25 \\
spectreconsole/spectre.console & 729 & 62.5k & 10 & 172.10 & 55.60 & 7.40 & 4.70 & 68.20 & 9.10 & 6.40 & 1.20 & 0.10 \\
\bottomrule
\end{tabular}
}
\end{center}
\end{table*}
\begin{table*}[!htb]
    \begin{center}
\captionof{table}{Statistics of SWE-Bench Verified}
\label{tab:instance_statistics_swe-bench-verified}
\resizebox{\textwidth}{!}{%
\begin{tabular}{l|rr|rr|rrr|rrr|rr}
\toprule
\multirow{2}{*}{\textbf{Org/Repo}} & \multicolumn{2}{c|}{\textbf{Repository}} & \multicolumn{2}{c|}{\textbf{Instance}} & \multicolumn{3}{c|}{\textbf{Gold Patch}} & \multicolumn{3}{c|}{\textbf{Test Patch}} & \multicolumn{2}{c}{\textbf{Unit tests}} \\
& \textbf{\#Files} & \textbf{\#LoC} & \textbf{\#Count} & \textbf{Avg. \#Tokens} & \textbf{Avg. \#Lines} & \textbf{Avg. \#Hunks} & \textbf{Avg. \#Files} & \textbf{Avg. \#Lines} & \textbf{Avg. \#Hunks} & \textbf{Avg. \#Files} & \textbf{\#Avg. F2P} & \textbf{\# Avg. P2P} \\
\midrule
astropy/astropy & 526 & 160.9k & 22 & 402.91 & 27.77 & 2.27 & 1.23 & 38.27 & 2.36 & 1.18 & 2.23 & 167.55 \\
django/django & 833 & 114.1k & 231 & 196.99 & 11.81 & 2.06 & 1.2 & 24.98 & 2.39 & 1.42 & 4.58 & 83.32 \\
matplotlib/matplotlib & 265 & 124.1k & 34 & 331.76 & 9.26 & 2.09 & 1.18 & 19.5 & 1.53 & 1.03 & 1.82 & 378.29 \\
mwaskom/seaborn & 63 & 23.4k & 2 & 237.0 & 13.5 & 3.0 & 1.5 & 18.5 & 1.5 & 1.5 & 2.0 & 171.0 \\
pallets/flask & 25 & 6.8k & 1 & 45.0 & 3.0 & 1.0 & 1.0 & 5.0 & 1.0 & 1.0 & 1.0 & 59.0 \\
psf/requests & 21 & 4.4k & 8 & 224.88 & 3.62 & 1.5 & 1.0 & 6.0 & 1.0 & 1.0 & 3.88 & 100.25 \\
pydata/xarray & 132 & 82.0k & 22 & 392.68 & 18.0 & 2.36 & 1.23 & 22.68 & 1.55 & 1.18 & 2.18 & 657.73 \\
pylint-dev/pylint & 1082 & 43.5k & 10 & 450.8 & 24.7 & 3.7 & 2.1 & 33.8 & 1.9 & 1.2 & 4.2 & 46.3 \\
pytest-dev/pytest & 85 & 28.5k & 19 & 336.42 & 22.58 & 2.42 & 1.11 & 48.32 & 3.11 & 1.26 & 1.84 & 64.74 \\
scikit-learn/scikit-learn & 423 & 179.7k & 32 & 379.81 & 12.31 & 2.97 & 1.06 & 20.22 & 1.72 & 1.06 & 1.22 & 64.19 \\
sphinx-doc/sphinx & 226 & 66.8k & 44 & 249.82 & 16.84 & 2.45 & 1.25 & 25.52 & 2.34 & 1.55 & 1.2 & 23.36 \\
sympy/sympy & 858 & 383.4k & 75 & 162.27 & 16.63 & 3.56 & 1.44 & 14.63 & 2.11 & 1.28 & 1.25 & 51.99 \\
\bottomrule
\end{tabular}
}
\end{center}
\end{table*}
\begin{table*}[!htb]
    \begin{center}
\captionof{table}{Statistics of Multi-SWE-Bench (Java only)}
\label{tab:multi_swe_bench_statistics}
\resizebox{\textwidth}{!}{%
\begin{tabular}{l|rr|rr|rrr|rrr}
\toprule
\multirow{2}{*}{\textbf{}} & \multicolumn{2}{c|}{\textbf{Repository}} & \multicolumn{2}{c|}{\textbf{Instance}} & \multicolumn{3}{c|}{\textbf{Fix patches}} & \multicolumn{3}{c}{\textbf{Unit tests}} \\
\textbf{Org/Repo} & \textbf{\#Files} & \textbf{\#LoC} & \textbf{\#Num} & \textbf{Avg. \#Tokens} & \textbf{Avg. \#Lines} & \textbf{Avg. \#Hunks} & \textbf{Avg. \#Files} & \textbf{\#A2P2P} & \textbf{\#A2F2P} & \textbf{\#A2N2P} \\
\midrule
\multicolumn{11}{c}{\textbf{Java}} \\
\midrule
alibaba/fastjson2 & 4244 & 443.8k & 6 & 459.2 & 10.5 & 1.3 & 1.2 & 1243.5 & 0.8 & 1020.5 \\
elastic/logstash & 562 & 59.9k & 38 & 1600.4 & 212.3 & 10.0 & 4.6 & 554.7 & 1.9 & 256.2 \\
mockito/mockito & 986 & 84.0k & 6 & 315.2 & 92.5 & 10.3 & 4.7 & 97.2 & 1.0 & 3.8 \\
apache/dubbo & 3939 & 402.1k & 3 & 774.0 & 9.3 & 3.0 & 1.3 & 2.0 & 57.0 & 0.0 \\
fasterxml/j-core & 366 & 105.7k & 18 & 304.7 & 33.8 & 4.8 & 2.1 & 2.0 & 85.6 & 0.0 \\
fasterxml/j-dbind & 1230 & 217.5k & 42 & 621.5 & 35.1 & 3.9 & 2.1 & 2.0 & 73.8 & 0.0 \\
fasterxml/j-dfmt-xml & 206 & 23.0k & 5 & 1071.8 & 98.4 & 10.4 & 3.2 & 2.0 & 94.2 & 0.0 \\
google/gson & 261 & 48.0k & 5 & 365.8 & 35.8 & 4.6 & 1.8 & 2.0 & 62.6 & 0.0 \\
google-ct/jib & 604 & 75.5k & 5 & 1094.6 & 15.2 & 3.2 & 2.6 & 2.0 & 96.2 & 0.0 \\
\midrule
\multicolumn{11}{c}{\textbf{TypeScript}} \\
\midrule
darkreader/darkreader & 189 & 26.2k & 2 & 749.5 & 13.0 & 2.0 & 1.5 & 41.0 & 3.5 & 0.0 \\
mui/material-ui & 27632 & 698.6k & 174 & 508.6 & 331.2 & 20.2 & 12.0 & 5001.3 & 2.3 & 836.8 \\
type/core & 509 & 128.2k & 48 & 694.8 & 22.9 & 3.5 & 1.9 & 2920.4 & 3.0 & 0.0 \\
\midrule
\multicolumn{11}{c}{\textbf{JavaScript}} \\
\midrule
ag/gh-rtdme-stats & 69 & 11.8k & 19 & 287.1 & 123.6 & 13.5 & 4.8 & 108.9 & 3.5 & 3.4 \\
axios/axios & 166 & 21.0k & 4 & 490.8 & 179.5 & 7.8 & 4.0 & 68.5 & 1.2 & 0.0 \\
expressjs/express & 142 & 17.3k & 4 & 177.5 & 7.2 & 2.2 & 1.5 & 808.2 & 1.5 & 65.2 \\
iamkun/dayjs & 324 & 17.1k & 56 & 325.6 & 21.7 & 2.7 & 2.0 & 60.4 & 1.2 & 3.2 \\
Kong/Insomnia & 526 & 182.0k & 1 & 709.0 & 1.0 & 1.0 & 1.0 & 105.0 & 1.0 & 0.0 \\
sveltejs/svelte & 2800 & 105.9k & 272 & 618.9 & 72.0 & 8.4 & 4.0 & 4904.2 & 5.5 & 0.0 \\
\midrule
\multicolumn{11}{c}{\textbf{Go}} \\
\midrule
cli/cli & 737 & 165.1k & 397 & 347.6 & 103.8 & 9.0 & 3.9 & 1997.0 & 2.9 & 31.0 \\
grpc/grpc-go & 981 & 260.8k & 16 & 276.1 & 81.8 & 7.7 & 2.8 & 230.4 & 0.6 & 6.6 \\
zeromicro/go-zero & 960 & 117.6k & 15 & 205.2 & 52.4 & 4.9 & 2.7 & 1318.9 & 0.3 & 43.9 \\
\midrule
\multicolumn{11}{c}{\textbf{Rust}} \\
\midrule
BurntSushi/ripgrep & 98 & 45.4k & 14 & 553.7 & 1604.9 & 21.9 & 7.5 & 233.2 & 1.1 & 8.1 \\
clap-rs/clap & 321 & 70.4k & 132 & 987.0 & 147.1 & 15.7 & 4.7 & 489.5 & 3.1 & 378.8 \\
nushell/nushell & 1479 & 264.2k & 14 & 795.6 & 155.0 & 10.6 & 4.3 & 798.6 & 2.6 & 336.6 \\
rayon-rs/rayon & 191 & 36.9k & 2 & 153.5 & 637.5 & 5.5 & 2.0 & 113.5 & 0.5 & 171.0 \\
serde-rs/serde & 188 & 36.5k & 2 & 171.5 & 72.5 & 3.0 & 3.0 & 0.0 & 0.0 & 294.5 \\
sharkdp/bat & 83 & 22.0k & 10 & 638.2 & 239.5 & 14.1 & 5.9 & 152.7 & 1.7 & 33.6 \\
sharkdp/fd & 24 & 6.7k & 14 & 167.8 & 55.8 & 7.8 & 4.5 & 186.5 & 1.1 & 0.0 \\
tokio-rs/bytes & 33 & 11.9k & 5 & 188.0 & 45.0 & 5.6 & 1.8 & 23.2 & 0.4 & 91.6 \\
tokio-rs/tokio & 727 & 141.5k & 25 & 590.0 & 139.8 & 10.6 & 3.5 & 26.6 & 0.0 & 287.4 \\
tokio-rs/tracing & 241 & 60.9k & 21 & 472.0 & 597.2 & 39.3 & 7.1 & 30.8 & 0.2 & 182.0 \\
\midrule
\multicolumn{11}{c}{\textbf{C}} \\
\midrule
facebook/zstd & 276 & 119.8k & 29 & 496.6 & 67.6 & 10.9 & 3.0 & 0.8 & 0.5 & 5.6 \\
jqlang/jq & 80 & 43.0k & 17 & 429.8 & 26.1 & 2.7 & 1.8 & 27.2 & 1.0 & 0.1 \\
ponylang/ponyc & 285 & 80.2k & 82 & 480.2 & 205.4 & 15.6 & 5.7 & 997.6 & 1.9 & 388.8 \\
\midrule
\multicolumn{11}{c}{\textbf{C++}} \\
\midrule
catchorg/Catch2 & 399 & 58.0k & 12 & 357.3 & 469.0 & 15.4 & 8.2 & 19.9 & 0.7 & 17.6 \\
fmtlib/fmt & 25 & 36.4k & 41 & 397.7 & 36.8 & 3.0 & 1.1 & 9.3 & 0.0 & 9.3 \\
nlohmann/json & 477 & 124.7k & 55 & 905.5 & 405.8 & 27.9 & 6.5 & 26.5 & 0.0 & 42.9 \\
simdutf/simdutf & 455 & 229.7k & 20 & 320.2 & 768.5 & 35.5 & 11.0 & 18.6 & 0.0 & 41.5 \\
yhirose/cpp-httplib & 33 & 50.9k & 1 & 240.0 & 1.0 & 1.0 & 1.0 & 272.0 & 1.0 & 0.0 \\
\bottomrule
\end{tabular}
}
\end{center}
\end{table*}
\clearpage
\subsection{Extended Results}
We use this section to provide some additional results which include performance of different models with respect to different dimensions. All the SWE-Sharp-Bench agent runs were scheduled by us, using the prompt template mentioned in Figure \ref{fig:agent_prompt}. Table \ref{tab:benchmark_results} provides resolution rate for model + agent configurations combinations across all three benchmarks. Resolution rates reported for SWE-Bench Verified and Multi-SWE-Bench are obtained from their respective public leader-boards \footnote{\url{https://www.swebench.com/}, \url{https://multi-swe-bench.github.io/}}. SWE-Agent + Claude 4 Sonnet is one exception to this, this was scheduled by us. Due to resource constraints we were only able to do a single entire benchmark with this configuration. 

\begin{figure*}[!ht]
\begin{mdframed}[linewidth=1pt, roundcorner=5pt]
\begin{lstlisting}[basicstyle=\ttfamily\small, breaklines=true]
You are a helpful assistant that can interact with a computer to solve tasks.

<uploaded_files>
{{working_dir}}
</uploaded_files>

I've uploaded a C# repository in the directory {{working_dir}}. Consider the following PR description:

<pr_description>
{{problem_statement}}
</pr_description>

Can you help me implement the necessary changes to the repository so that the requirements specified in the <pr_description> are met?
I've already taken care of all changes to any of the test files described in the <pr_description>. This means you DON'T have to modify the testing logic or any of the tests in any way!
Your task is to make the minimal changes to non-tests files in the {{working_dir}} directory to ensure the <pr_description> is satisfied.
Follow these steps to resolve the issue:
1. As a first step, it might be a good idea to find and read code relevant to the <pr_description>
2. Create a script to reproduce the error and execute it using the bash tool, to confirm the error
3. Edit the sourcecode of the repo to resolve the issue
4. Rerun your reproduce script and confirm that the error is fixed!
5. Think about edgecases and make sure your fix handles them as well
Your thinking should be thorough and so it's fine if it's very long.
\end{lstlisting}
\caption{Prompt Template used for SWE-Agent and OpenHands run on SWE-Sharp-Bench}
\label{fig:agent_prompt}
\end{mdframed}
\end{figure*}

\begin{table*}
\centering
\caption{AI Agent Benchmark Resolution Rates}
\label{tab:benchmark_results}
\begin{tabular}{|l|l|r|r|r|}
\hline
\textbf{Agent} & \textbf{Model} & \multicolumn{3}{c|}{\textbf{Resolution Rate (\%)}} \\
\cline{3-5}
 &  & \textbf{Python - SWE-Bench Verified} & \textbf{Java - Multi-SWE-Bench} & \textbf{C\# - SWE-Sharp-Bench} \\
\hline
\multirow{4}{*}{\textbf{SWE-agent}} & GPT-4o & 26.00 & 5.11 & 11.30 \\
 & Claude Sonnet 3.5 & 33.60 & 10.20 & 20.00 \\
 & Claude Sonnet 3.7 & 62.40 & 14.68 & 31.30 \\
 & Claude Sonnet 4   & 66.60 & 18.75*  & 44.70 \\
\hline
\multirow{7}{*}{\textbf{OpenHands}} & GPT-4o & 25.75 & 5.96 & 8.00 \\
 & GPT-4.1 & 48.60 & 10.11 & 23.30 \\
 & OpenAI o3-mini & 43.70 & 6.29 & 19.33 \\
 & OpenAI o3 & 59.00 & 21.00 & 35.00 \\
 & Claude Sonnet 3.5 & 53.00 & 12.73 & 22.60 \\
 & Claude Sonnet 3.7 & 60.60 & 16.01 & 31.30 \\
\hline
\end{tabular}
\end{table*}

\clearpage
\subsubsection{\textbf{Performance by Repository: }}\label{app:repo_perf}
In Table \ref{tab:repo_comparison} we provide a breakdown of repository-level performance across 2 dimensions: Agent and Model. We used GPT-4o and GPT-5 from OpenAI, and Claude Sonnet 3.5 and Claude Sonnet 4 from Anthropic. We use these combinations to show progress between model generations. 

\begin{table}[H]
\caption{Repository-wise Resolve Rate Comparison}
\label{tab:repo_comparison}
\small
\begin{adjustbox}{width=\textwidth,center}
\begin{tabular}{l|cccc|cccc}
\toprule
& \multicolumn{4}{c|}{\textbf{OpenHands}} & \multicolumn{4}{c}{\textbf{SWE-Agent}} \\
\cmidrule(lr){2-5} \cmidrule(lr){6-9}
\textbf{Repository} & \textbf{GPT-4o} & \textbf{GPT-5} & \textbf{Sonnet-3.5} & \textbf{Sonnet-4} & \textbf{GPT-4o} & \textbf{GPT-5} & \textbf{Sonnet-3.5} & \textbf{Sonnet-4} \\
\midrule
dotnet/efcore & 0.0 & 55.6 & 33.3 & 44.4 & 55.6 & 55.6 & 44.4 & 55.6 \\
serilog/serilog & 33.3 & 50.0 & 33.3 & 58.3 & 16.7 & 41.7 & 33.3 & 58.3 \\
spectreconsole/spectre.console & 10.0 & 60.0 & 40.0 & 60.0 & 10.0 & 60.0 & 20.0 & 60.0 \\
SixLabors/ImageSharp & 0.0 & 62.5 & 25.0 & 75.0 & 12.5 & 62.5 & 25.0 & 50.0 \\
autofac/Autofac & 16.7 & 50.0 & 33.3 & 50.0 & 0.0 & 50.0 & 33.3 & 50.0 \\
restsharp/RestSharp & 0.0 & 60.0 & 20.0 & 60.0 & 0.0 & 60.0 & 20.0 & 60.0 \\
dotnet/BenchmarkDotNet & 12.5 & 62.5 & 25.0 & 50.0 & 0.0 & 62.5 & 12.5 & 50.0 \\
devlooped/moq & 0.0 & 50.0 & 25.0 & 75.0 & 0.0 & 50.0 & 0.0 & 75.0 \\
AvaloniaUI/Avalonia & 9.8 & 41.5 & 24.4 & 31.7 & 14.6 & 36.6 & 22.0 & 41.5 \\
StackExchange/StackExchange.Redis & 0.0 & 33.3 & 0.0 & 33.3 & 0.0 & 33.3 & 33.3 & 66.7 \\
App-vNext/Polly & 0.0 & 42.9 & 21.4 & 28.6 & 7.1 & 42.9 & 14.3 & 35.7 \\
MessagePack-CSharp/MessagePack-CSharp & 0.0 & 50.0 & 5.6 & 27.8 & 5.6 & 44.4 & 5.6 & 38.9 \\
ardalis/CleanArchitecture & 33.3 & 33.3 & 33.3 & 0.0 & 0.0 & 33.3 & 33.3 & 0.0 \\
ThreeMammals/Ocelot & 0.0 & 25.0 & 0.0 & 50.0 & 0.0 & 0.0 & 0.0 & 0.0 \\
jellyfin/jellyfin & 0.0 & 33.3 & 0.0 & 0.0 & 0.0 & 0.0 & 0.0 & 33.3 \\
gui-cs/Terminal.Gui & 0.0 & 0.0 & 0.0 & 0.0 & 0.0 & 0.0 & 0.0 & 0.0 \\
JoshClose/CsvHelper & 0.0 & 0.0 & 0.0 & 0.0 & 0.0 & 0.0 & 0.0 & 0.0 \\
\midrule
\textbf{OVERALL} & \textbf{8.0} & \textbf{47.3} & \textbf{22.6} & \textbf{40.6} & \textbf{11.3} & \textbf{43.3} & \textbf{20.0} & \textbf{44.7} \\
\bottomrule
\end{tabular}
\end{adjustbox}
\end{table}

\subsubsection{\textbf{Performance by Issue Creation Year:}}
Table \ref{tab:year_comparison} provides a temporal breakdown that shows resolution for instances across different years. We use the same model + agent configurations mentioned in \ref{app:repo_perf}. 

\begin{table}[H]
    \centering
\caption{Year-wise Resolve Rate Comparison}
\label{tab:year_comparison}
\small
\begin{tabular}{c|cccc|cccc}
\toprule
& \multicolumn{4}{c|}{\textbf{OpenHands}} & \multicolumn{4}{c}{\textbf{SWE-Agent}} \\
\cmidrule(lr){2-5} \cmidrule(lr){6-9}
\textbf{Year} & \textbf{GPT-4o} & \textbf{GPT-5} & \textbf{Sonnet-3.5} & \textbf{Sonnet-4} & \textbf{GPT-4o} & \textbf{GPT-5} & \textbf{Sonnet-3.5} & \textbf{Sonnet-4} \\
\midrule
2020 & 0.0 & 60.0 & 20.0 & 80.0 & 0.0 & 60.0 & 0.0 & 80.0 \\
2021 & 0.0 & 100.0 & 100.0 & 100.0 & 0.0 & 100.0 & 100.0 & 100.0 \\
2022 & 0.0 & 44.4 & 22.2 & 55.6 & 33.3 & 44.4 & 22.2 & 33.3 \\
2023 & 7.4 & 42.6 & 20.4 & 40.7 & 3.7 & 38.9 & 18.5 & 38.9 \\
2024 & 10.4 & 50.6 & 23.4 & 36.4 & 15.6 & 45.5 & 22.1 & 48.1 \\
2025 & 0.0 & 25.0 & 25.0 & 25.0 & 0.0 & 25.0 & 0.0 & 25.0 \\
\midrule
\textbf{TOTAL} & \textbf{8.0} & \textbf{47.3} & \textbf{22.6} & \textbf{40.6} & \textbf{11.3} & \textbf{43.3} & \textbf{20.0} & \textbf{44.7} \\
\bottomrule
\end{tabular}
\end{table}

\subsubsection{\textbf{Trajectory Analysis:}} 

\begin{figure*}[ht]
\centering

\begin{subfigure}{0.48\textwidth}
    \includegraphics[width=\textwidth,height=0.2\textheight,keepaspectratio]{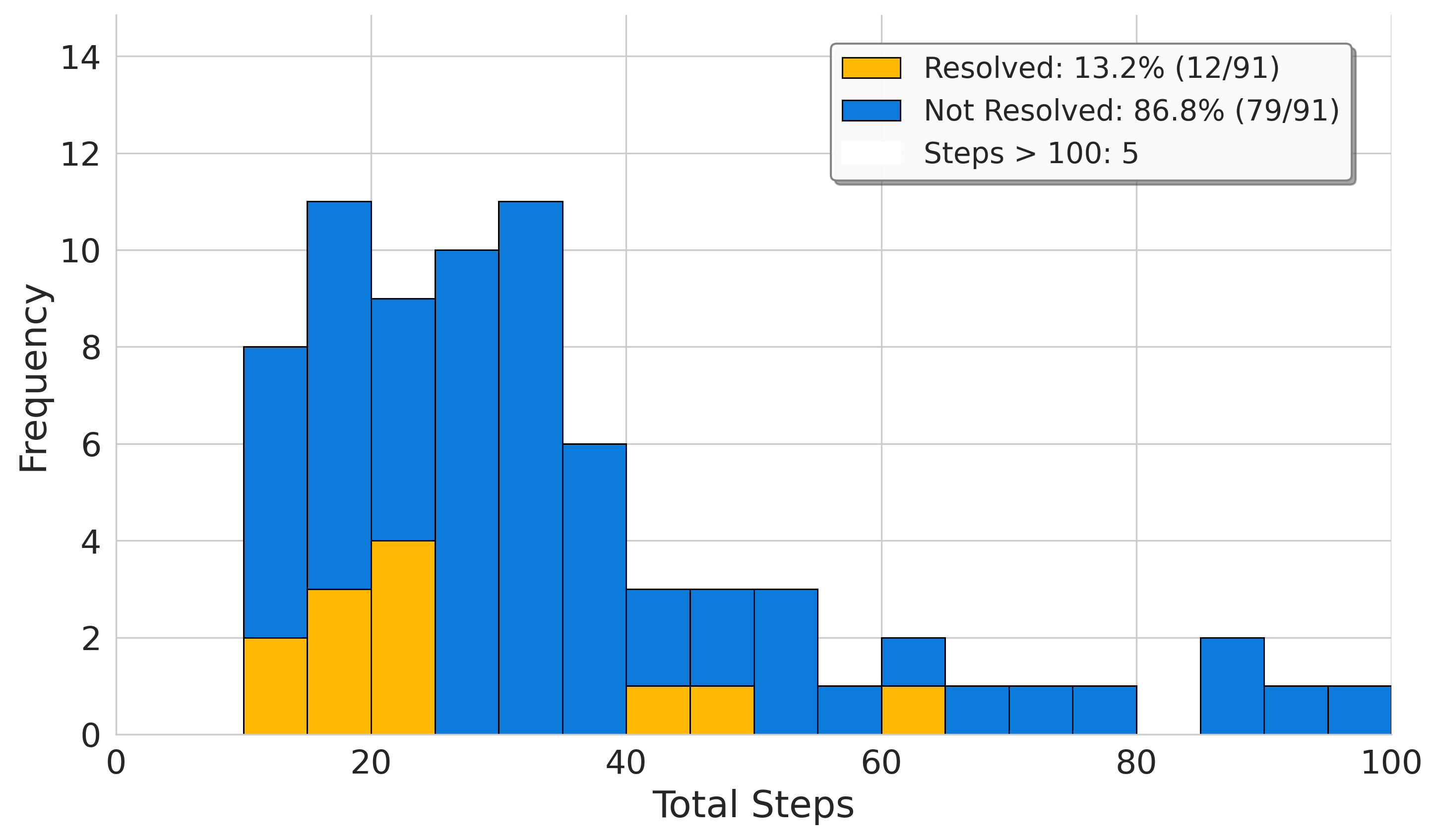}
    \caption{OpenHands + GPT-4o}
    \label{fig:oh_gpt4o}
\end{subfigure}
\hfill
\begin{subfigure}{0.48\textwidth}
    \includegraphics[width=\textwidth,height=0.2\textheight,keepaspectratio]{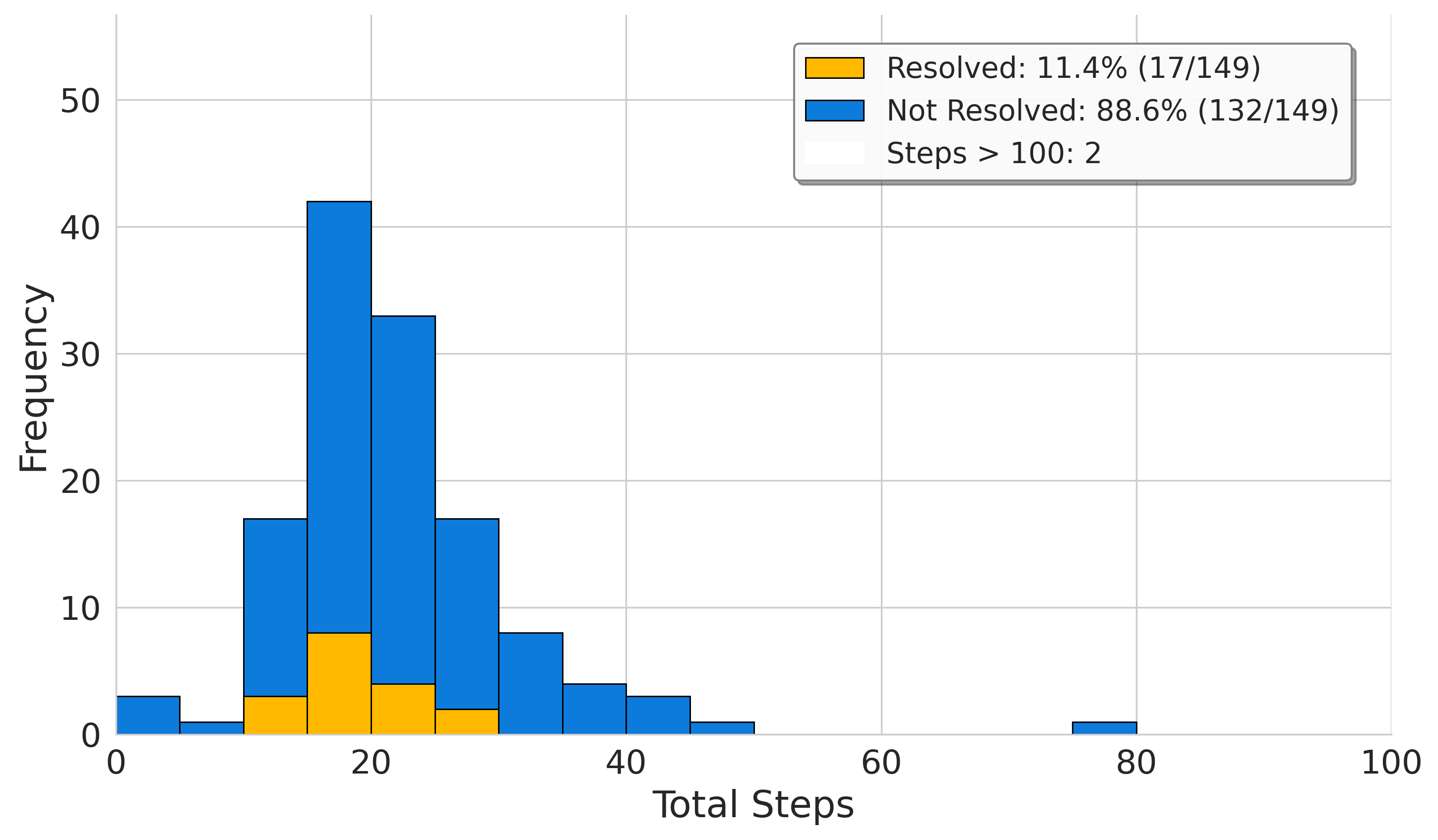}
    \caption{SWE-Agent + GPT-4o}
    \label{fig:swe_gpt4o}
\end{subfigure}

\begin{subfigure}{0.48\textwidth}
    \includegraphics[width=\textwidth,height=0.2\textheight,keepaspectratio]{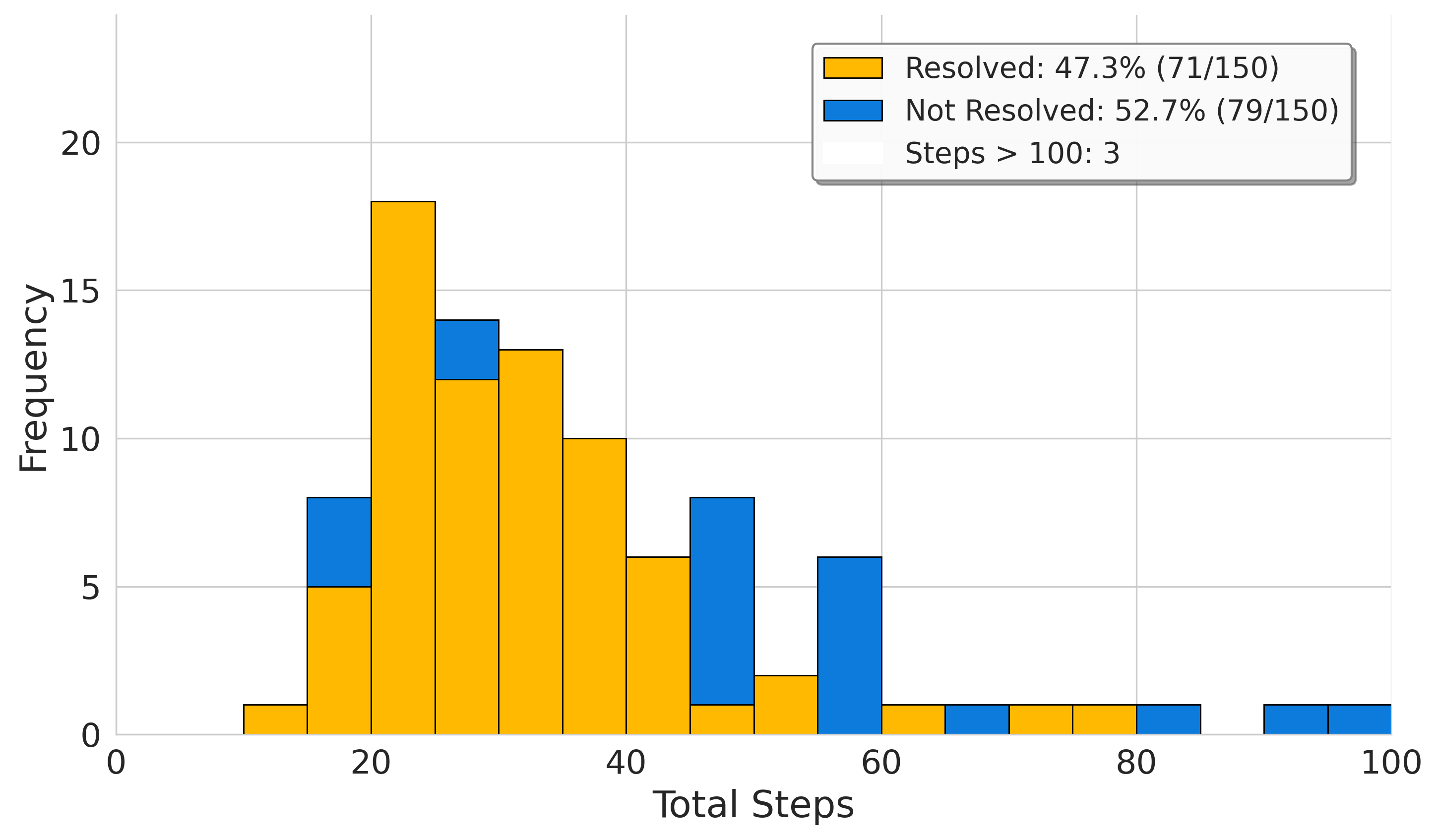}
    \caption{OpenHands + GPT-5}
    \label{fig:oh_gpt5}
\end{subfigure}
\hfill
\begin{subfigure}{0.48\textwidth}
    \includegraphics[width=\textwidth,height=0.2\textheight,keepaspectratio]{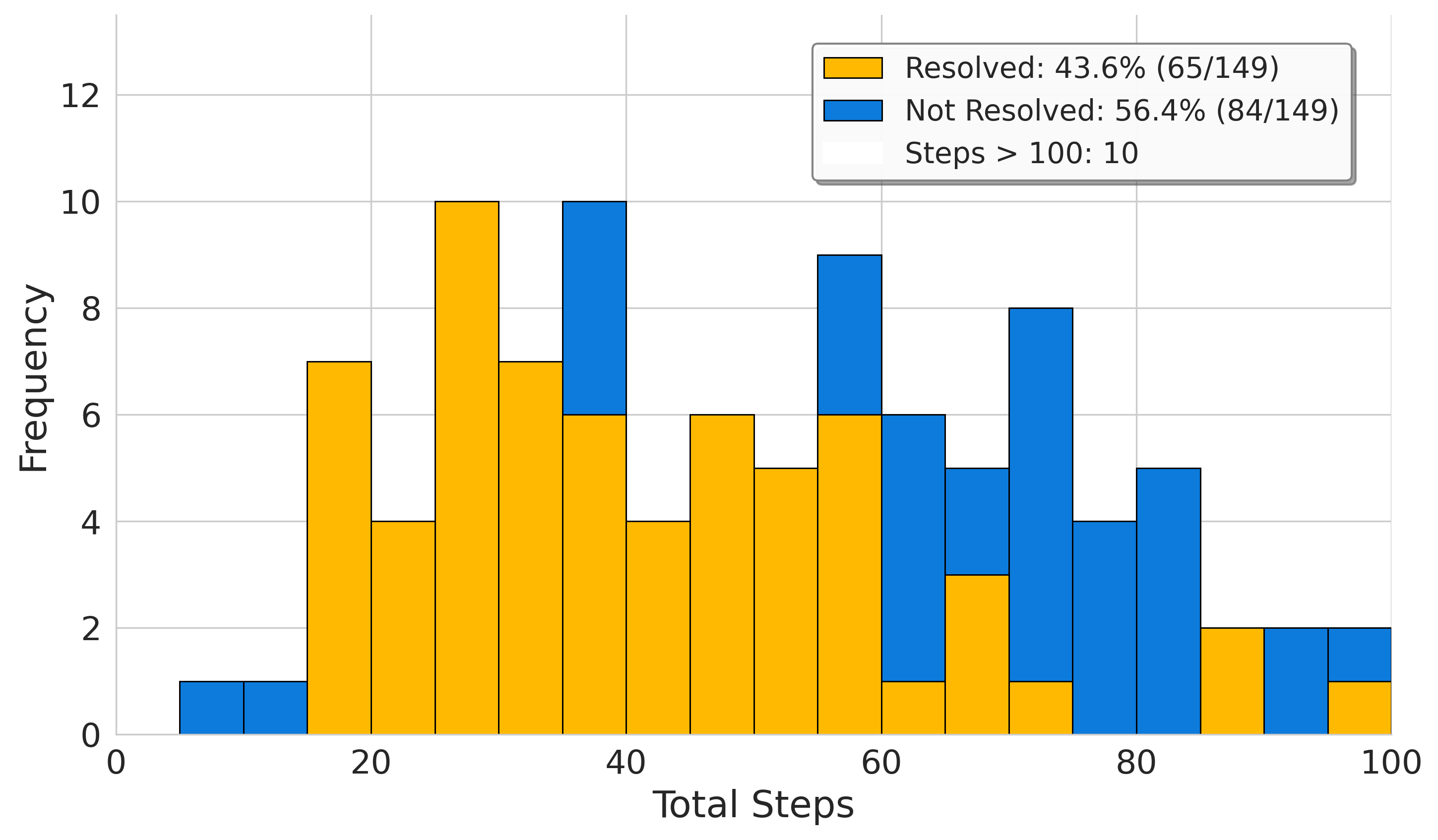}
    \caption{SWE-Agent + GPT-5}
    \label{fig:swe_gpt5}
\end{subfigure}

\begin{subfigure}{0.48\textwidth}
    \includegraphics[width=\textwidth,height=0.2\textheight,keepaspectratio]{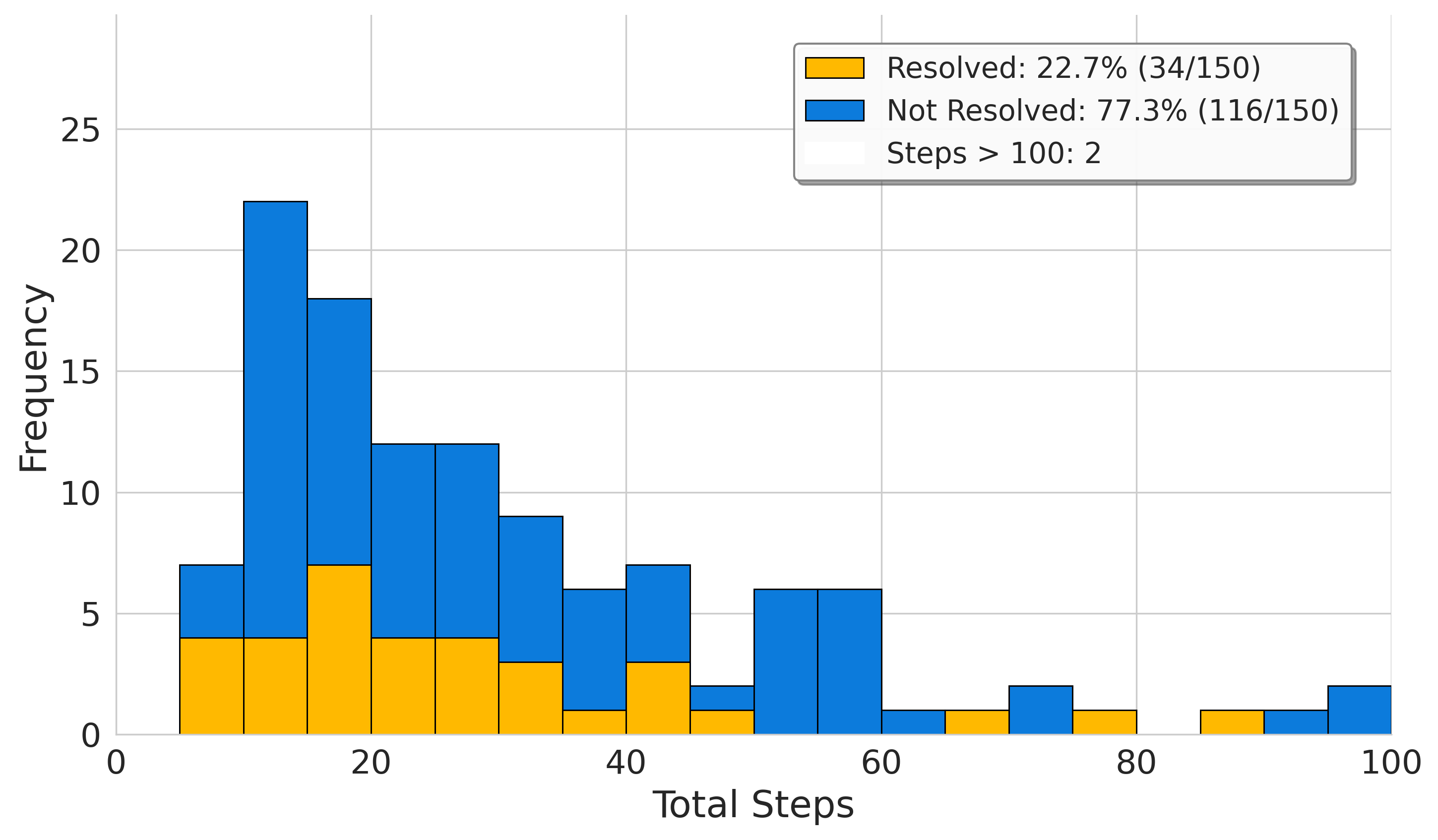}
    \caption{OpenHands + Sonnet 3.5}
    \label{fig:oh_sonnet35}
\end{subfigure}
\hfill
\begin{subfigure}{0.48\textwidth}
    \includegraphics[width=\textwidth,height=0.2\textheight,keepaspectratio]{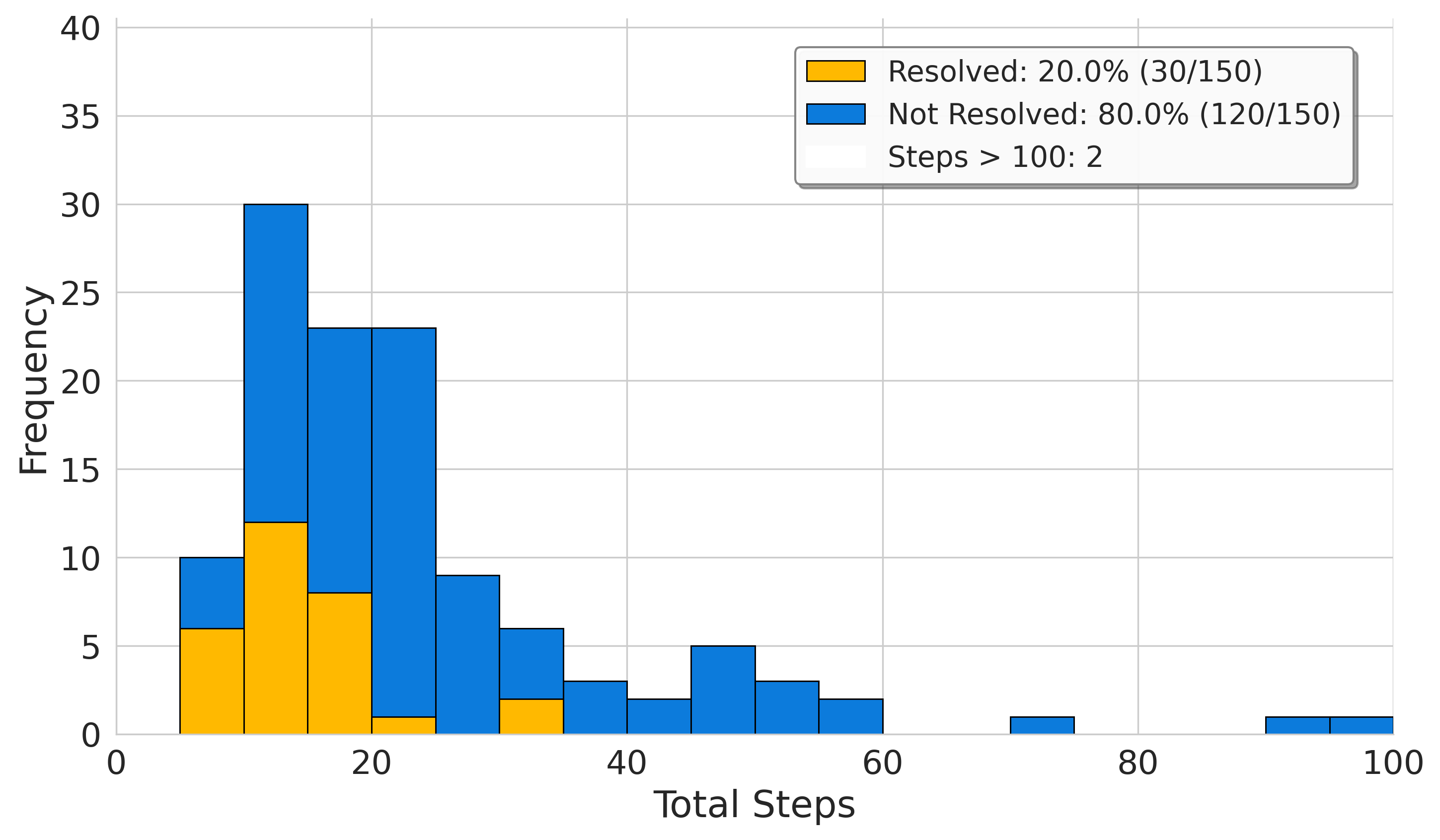}
    \caption{SWE-Agent + Sonnet 3.5}
    \label{fig:swe_sonnet35}
\end{subfigure}

\begin{subfigure}{0.48\textwidth}
    \includegraphics[width=\textwidth,height=0.2\textheight,keepaspectratio]{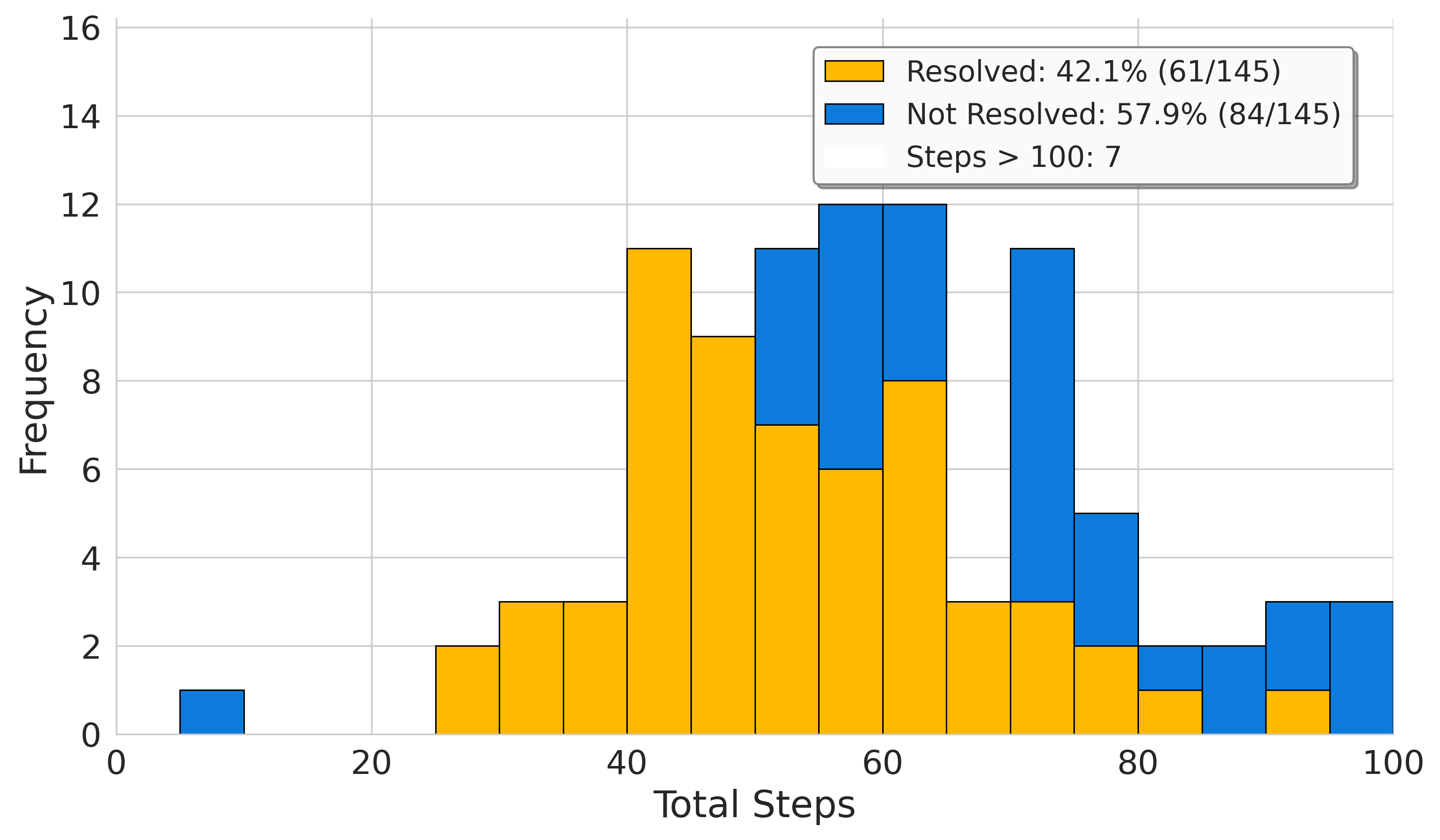}
    \caption{OpenHands + Sonnet 4}
    \label{fig:oh_sonnet4}
\end{subfigure}
\hfill
\begin{subfigure}{0.48\textwidth}
    \includegraphics[width=\textwidth,height=0.2\textheight,keepaspectratio]{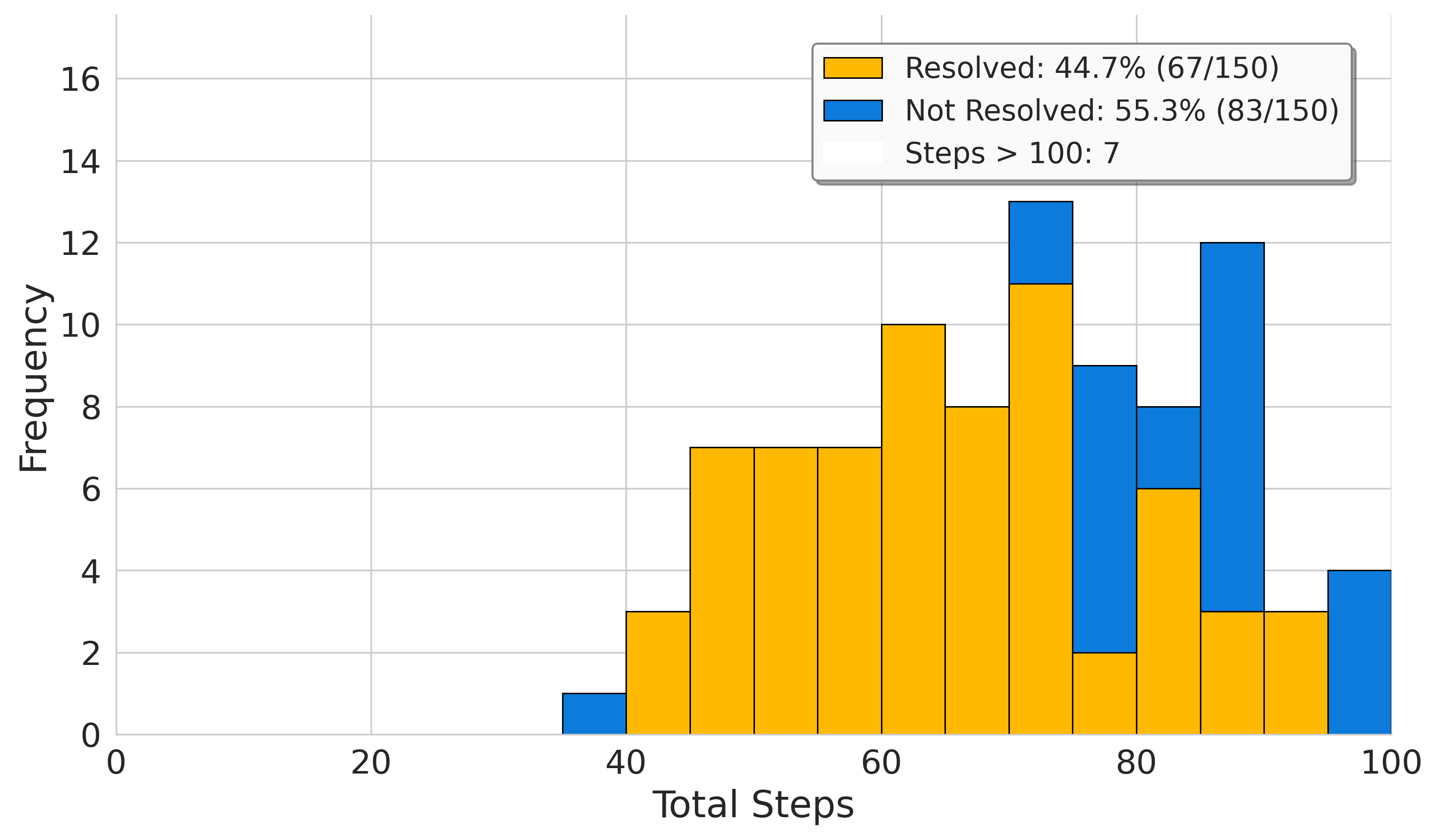}
    \caption{SWE-Agent + Sonnet 4}
    \label{fig:swe_sonnet4}
\end{subfigure}

\caption{Comparison of OpenHands vs SWE-Agent performance across different language models on SWE-Sharp-Bench}
\label{fig:turns_distribution}
\end{figure*}

\begin{figure*}[ht]
\centering

\begin{subfigure}{0.48\textwidth}
    \includegraphics[width=\textwidth,height=0.2\textheight,keepaspectratio]{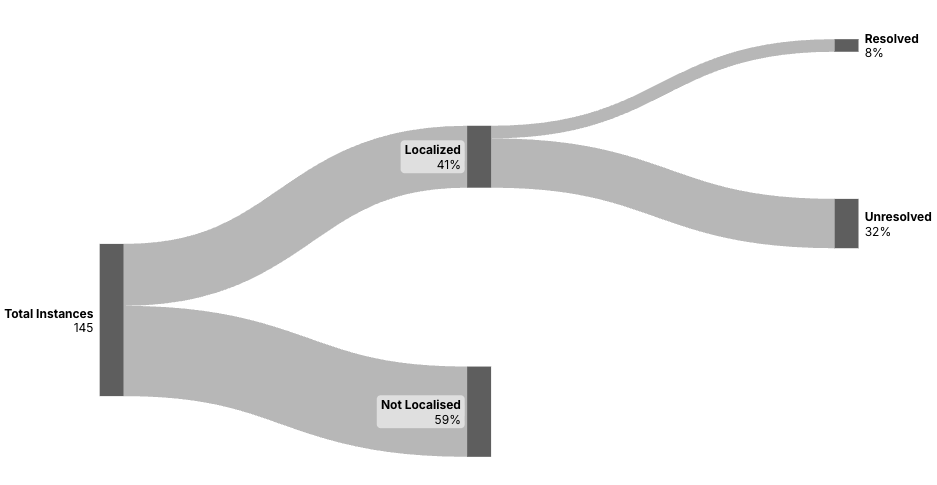}
    \caption{OpenHands + GPT-4o}
    \label{fig:oh_gpt4o}
\end{subfigure}
\hfill
\begin{subfigure}{0.48\textwidth}
    \includegraphics[width=\textwidth,height=0.2\textheight,keepaspectratio]{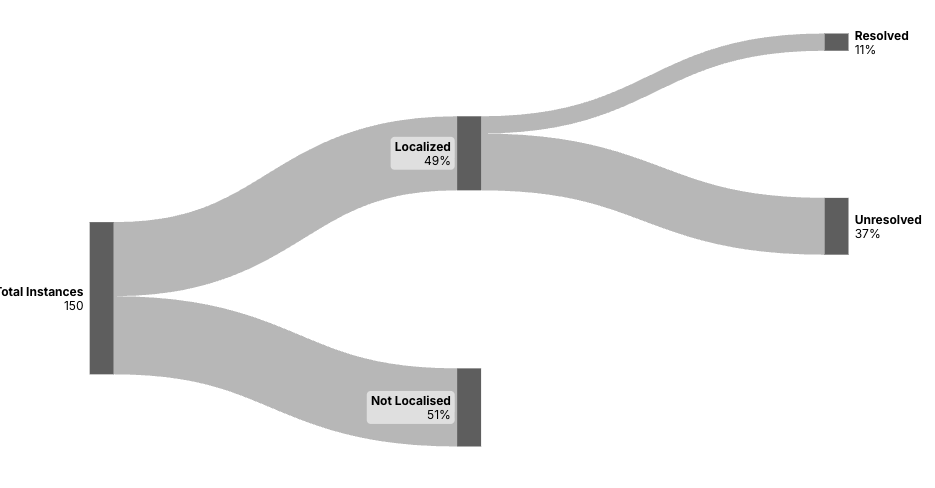}
    \caption{SWE-Agent + GPT-4o}
    \label{fig:swe_gpt4o}
\end{subfigure}

\begin{subfigure}{0.48\textwidth}
    \includegraphics[width=\textwidth,height=0.2\textheight,keepaspectratio]{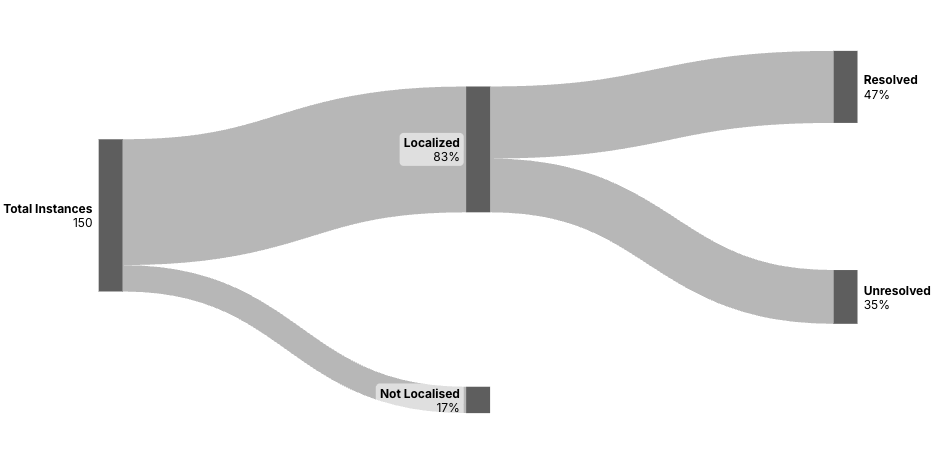}
    \caption{OpenHands + GPT-5}
    \label{fig:oh_gpt5}
\end{subfigure}
\hfill
\begin{subfigure}{0.48\textwidth}
    \includegraphics[width=\textwidth,height=0.2\textheight,keepaspectratio]{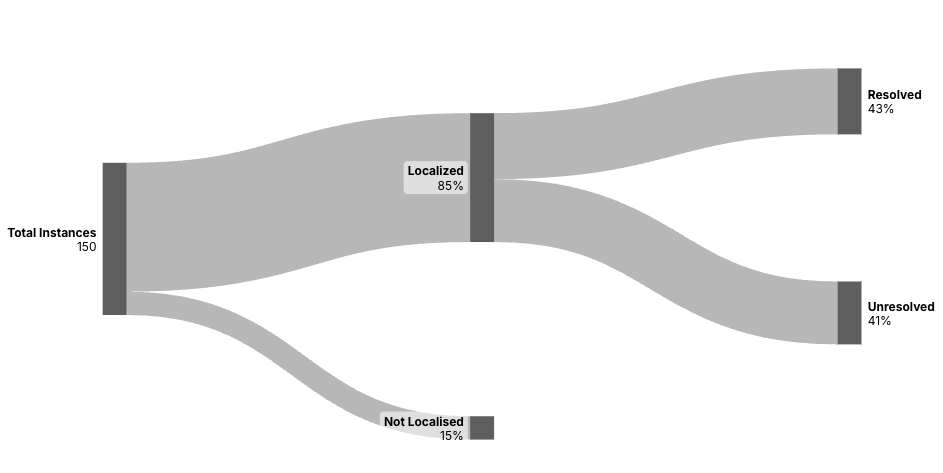}
    \caption{SWE-Agent + GPT-5}
    \label{fig:swe_gpt5}
\end{subfigure}

\begin{subfigure}{0.48\textwidth}
    \includegraphics[width=\textwidth,height=0.2\textheight,keepaspectratio]{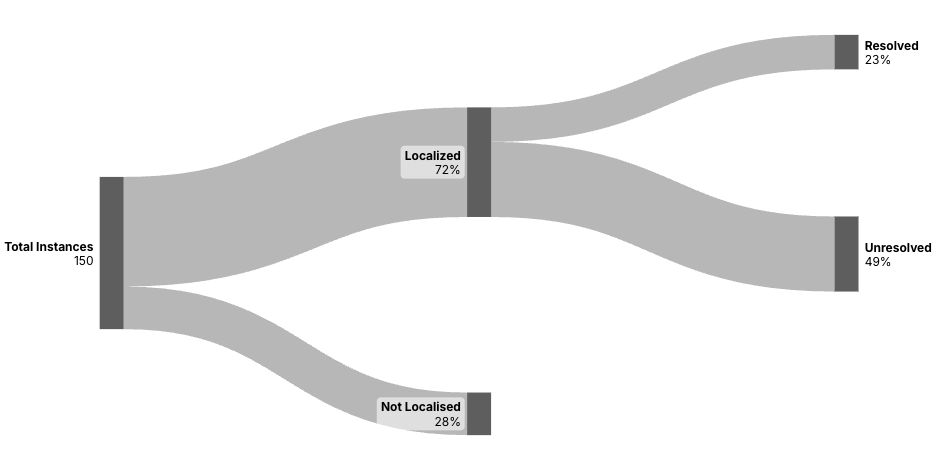}
    \caption{OpenHands + Sonnet 3.5}
    \label{fig:oh_sonnet35}
\end{subfigure}
\hfill
\begin{subfigure}{0.48\textwidth}
    \includegraphics[width=\textwidth,height=0.2\textheight,keepaspectratio]{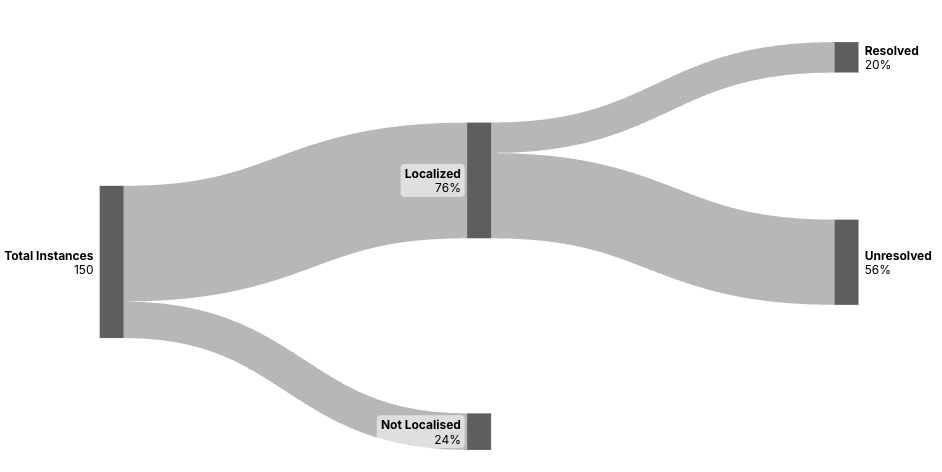}
    \caption{SWE-Agent + Sonnet 3.5}
    \label{fig:swe_sonnet35}
\end{subfigure}

\begin{subfigure}{0.48\textwidth}
    \includegraphics[width=\textwidth,height=0.2\textheight,keepaspectratio]{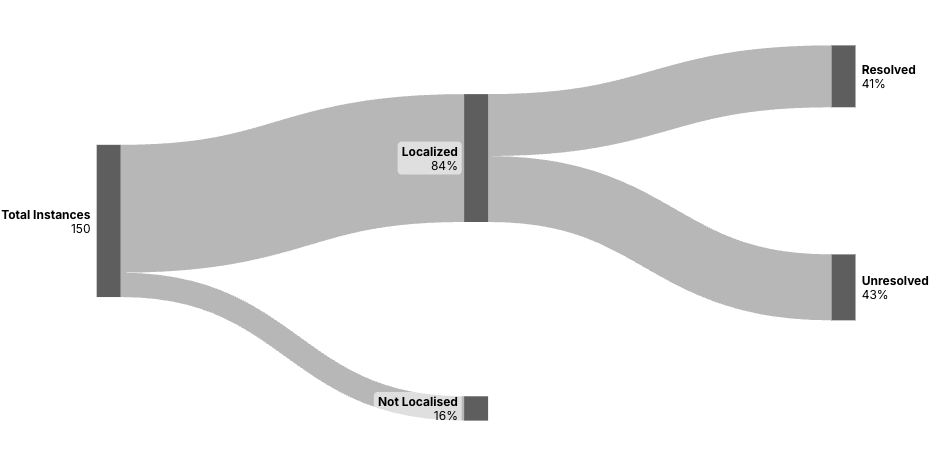}
    \caption{OpenHands + Sonnet 4}
    \label{fig:oh_sonnet4}
\end{subfigure}
\hfill
\begin{subfigure}{0.48\textwidth}
    \includegraphics[width=\textwidth,height=0.2\textheight,keepaspectratio]{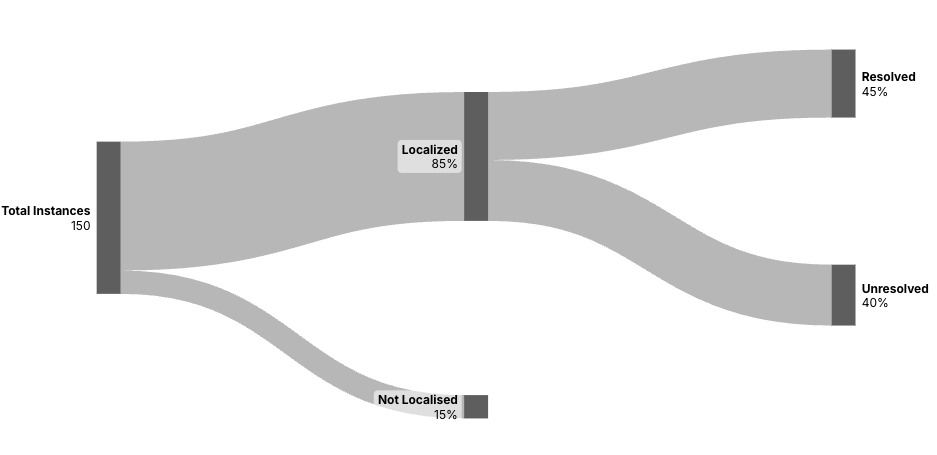}
    \caption{SWE-Agent + Sonnet 4}
    \label{fig:swe_sonnet4}
\end{subfigure}

\caption{Localization to Resolution flow }
\label{fig:localization_flow}
\end{figure*}

In this section we provide trajectory analysis of the same model + agent combinations from the previous sections. Figure \ref{fig:turns_distribution} visualizes the distribution of the number of turns that were resolved. Figure \ref{fig:localization_flow} demonstrates the percentage of instances that were successfully localized by the agent and resolved finally. The criteria of successful localization is the agent lands up at least one of the files from the golden fix patch.

\clearpage

\end{document}